\documentclass{jfm}
\usepackage{graphics}
\usepackage{epstopdf, epsfig}
\usepackage{upmath}

\usepackage{mathtools}
\DeclareFontFamily{U}{matha}{\hyphenchar\font45}
\DeclareFontShape{U}{matha}{m}{n}{ <-6> matha5 <6-7> matha6 <7-8>
matha7 <8-9> matha8 <9-10> matha9 <10-12> matha10 <12-> matha12 }{}
\DeclareSymbolFont{matha}{U}{matha}{m}{n}
\DeclareFontFamily{U}{mathx}{\hyphenchar\font45}
\DeclareFontShape{U}{mathx}{m}{n}{ <-6> mathx5 <6-7> mathx6 <7-8>
mathx7 <8-9> mathx8 <9-10> mathx9 <10-12> mathx10 <12-> mathx12 }{}
\DeclareSymbolFont{mathx}{U}{mathx}{m}{n}
 
\DeclareMathDelimiter{\ldbrack} {4}{matha}{"76}{mathx}{"30}
\DeclareMathDelimiter{\rdbrack} {5}{matha}{"77}{mathx}{"38}

\usepackage{multiJFMequations}
\usepackage{amsmath,bm}
\usepackage{color}
\newcommand*{\be}{\begin{equation}}
\newcommand*{\ee}{\end{equation}}
\newcommand*{\bse}{\begin{subequations}}
\newcommand*{\ese}{\end{subequations}}
\newcommand*{\bme}{\begin{multiequations}}
\newcommand*{\eme}{\end{multiequations}}
\newcommand*{\se}{\singleequation}
\newcommand*{\de}{\doubleequation}
\newcommand*{\te}{\tripleequation}

\newcommand*{\XXint}[3]{{\setbox0=\hbox{$#1{#2#3}{\int}$}
\vcenter{\hbox{$#2#3$}}\kern-.5\wd0}}

\def\pmsb#1{\setbox0=\hbox{#1}%
  \kern-.016em\copy0\kern-\wd0
  \kern.033em\copy0\kern-\wd0
  \kern-.016em\raise.0288em\box0}
\newcommand*{\btriangle}{\setbox0=\hbox{$\triangle$}%
  \kern-.016em\copy0\kern-\wd0
  \kern.033em\copy0\kern-\wd0
  \kern-.016em\raise.0288em\box0}
\newcommand*{\Bbtriangle}{\setbox0=\hbox{$\triangle$}%
  \kern-.02em\copy0\kern-\wd0
  \kern.033em\copy0\kern-\wd0
  \kern-.016em\raise.0144em\box0}
\newcommand*{\Btriangle}{\setbox0=\hbox{$\triangle$}%
  \kern-.025em\copy0\kern-\wd0
  \kern.05em\copy0\kern-\wd0
  \kern-.025em\raise.0433em\box0}
\newcommand*{\DeltaB}{{\pmb{\triangle}}}

\newcommand*{\XXnotinfty}[3]{{\setbox0=\hbox{$#1{#2#3}{\to}$}
\vcenter{\hbox{$#2#3$}}\kern-.5\wd0}}

\newcommand*{\ds}{\displaystyle}
\providecommand*{\dfrac}[2]{\ds\frac{#1}{#2}}

\renewcommand*{\tilde}{\widetilde}
\renewcommand*{\hat}{\widehat}
\renewcommand*{\bar}{\overline}
\newcommand*{\tint}{{\textstyle\int}}
\newcommand*\bfcdot{\bm{\cdot}}
\newcommand*\bfnabla{\bm{\nabla}}
\renewcommand*{\sp}[2]{{#1}\bfcdot{#2}}
\newcommand*{\vp}[2]{{#1}\times{#2}}

\newcommand*{\pdt}{{\partial_t}}\newcommand*{\pdr}{{\partial_r}}
\newcommand*{\pdz}{{\partial_z}}\newcommand*{\pdzii}{{\partial^2_z}}\newcommand*{\pdziv}{{\partial^4_z}}
\newcommand*{\tRR}{{\mbox{\tiny {\itshape R}}}}
\newcommand*{\pdR}{{\partial_\tRR\,}}
\newcommand*{\DSplus}{{\partial_{\tRR}^+}}\newcommand*{\DSminus}{{\partial_{\tRR}^-}}
\newcommand*{\tTT}{{\mbox{\tiny {\itshape T}}}}
\newcommand*{\pdT}{{\partial_\tTT\,}}
\newcommand*{\DST}{{{\sf D}_{\;\!\!\tTT\,}}}

\newcommand*{\oav}[1]{{\langle{#1}\rangle}}
\newcommand*{\bav}[1]{{\bigl\langle{#1}\bigr\rangle}}

\newcommand*{\oavd}[1]{{\oav{\hskip -0.7mm \oav{#1}\hskip -0.7mm}}}
\newcommand*{\bavd}[1]{{\bav{\hskip -0.8mm \bav{#1}\hskip -0.8mm}}}

\newcommand*{\bint}[3]{{\bigl\langle{#1}\bigr\rangle_{\!{#2}}^{\!{#3}}}}

\newcommand*{\bjump}[1]{{\bigl\ldbrack{#1}\bigr\rdbrack}}

\renewcommand*{\Lambda}{\varLambda}
\renewcommand*{\Upsilon}{\varUpsilon}
\renewcommand*{\Phi}{\varPhi}
\renewcommand*{\Psi}{\varPsi}
\renewcommand*{\Omega}{\varOmega}
\renewcommand*{\Theta}{\varTheta}
\renewcommand*{\Xi}{\varXi}

\newcommand*{\aR}{{\mathrm a}}\newcommand*{\dR}{{\mathrm d}}\newcommand*{\hR}{{\mathrm h}}\newcommand*{\iR}{{\mathrm i}}\newcommand*{\tR}{{\mathrm t}}
\newcommand*{\DR}{{\mathrm D}}\newcommand*{\JR}{{\mathrm J}}\newcommand*{\RR}{{\mathrm R}}\newcommand*{\TR}{{\mathrm T}}
\newcommand*{\Ra}{{\RR\aR}}\newcommand*{\Ta}{{\TR\aR}}
\newcommand*{\gv}{{\bm{g}}}\newcommand*{\nv}{{\bm{n}}}\newcommand*{\uv}{{\bm{u}}}\newcommand*{\zv}{{\bm{z}}}
\newcommand*{\Gv}{{\bm{G}}}
\newcommand*{\fS}{{\sf{f}}}\newcommand*{\gS}{{\sf{g}}}\newcommand*{\tS}{{\sf{t}}}
\newcommand*{\AS}{{\sf{A}}}\newcommand*{\GS}{{\sf{G}}}\newcommand*{\QS}{{\sf{Q}}}\newcommand*{\RS}{{\sf{R}}}\newcommand*{\TS}{{\sf{T}}}\newcommand*{\YS}{{\sf{Y}}}\newcommand*{\ZS}{{\sf{Z}}}
\newcommand*{\DC}{{\mathcal D}}\newcommand*{\FC}{{\mathcal F}}\newcommand*{\GC}{{\mathcal G}}\newcommand*{\HC}{{\mathcal H}}\newcommand*{\IC}{{\mathcal I}}\newcommand*{\JC}{{\mathcal J}}\newcommand*{\KC}{{\mathcal K}}\newcommand*{\LC}{{\mathcal L}}\newcommand*{\NC}{{\mathcal N}}\newcommand*{\PC}{{\mathcal P}}\newcommand*{\QC}{{\mathcal Q}}\newcommand*{\RC}{{\mathcal R}}\newcommand*{\UC}{{\mathcal U}}\newcommand*{\VC}{{\mathcal V}}\newcommand*{\WC}{{\mathcal W}}\newcommand*{\YC}{{\mathcal Y}}\newcommand*{\ZC}{{\mathcal Z}}
\newcommand*{\eG}{{\mathfrak e}}

\newcommand*{\nvh}{{\hat \nv}}\newcommand*{\zvh}{{\hat \zv}}

\newcommand*{\tA}{{\mbox{\tiny {\itshape A}}}}
\newcommand*{\tD}{{\mbox{\tiny {\itshape D}}}}
\newcommand*{\tE}{{\mbox{\tiny {\itshape E}}}}

\newcommand*{\tH}{{\mbox{\tiny {\itshape H}}}}
\newcommand*{\tM}{{\mbox{\tiny {\itshape M}}}}

\newcommand*{\tQ}{{\mbox{\tiny {\itshape Q}}}}
\newcommand*{\tX}{{\mbox{\tiny {\itshape X}}}}
\newcommand*{\tY}{{\mbox{\tiny {\itshape Y}}}}
\newcommand*{\tT}{{\!\!\:\mbox{\tiny {\itshape T}}}}
\newcommand*{\tW}{{\mbox{\tiny {\itshape W}}}}
\newcommand*{\tPP}{{\mbox{\tiny {\itshape PP}}}}

\newcommand*{\tWW}{{\mbox{\tiny {\itshape WW}}}}

\newcommand*{\tDNS}{{\mbox{\tiny {\itshape DNS}}}}

\newcommand*{\bsquare}{{\,{\bm \square}\,}}

\def\today{
\number\day\space
\ifcase\month\or
January\or February\or March\or April\or May\or June\or
July\or August\or September\or October\or November\or December\fi
\space\number\year}

\shorttitle{Slowly rotating B\'enard convection}
  
\shortauthor{A.~M.~Soward, L.~Oruba  and E.~Dormy}

\title{B\'enard convection in a slowly rotating penny shaped cylinder subject to constant heat flux boundary conditions}

\author{A. M. Soward\aff{1}\corresp{\email{andrew.soward@ncl.ac.uk, ludivine.oruba@latmos.ipsl.fr, Emmanuel.Dormy@ens.fr}},
  L. Oruba\aff{2}$\dag$
 \and E. Dormy\aff{3}$\dag$}

\affiliation{\aff{1} School of Mathematics, Statistics and Physics, Newcastle University, Newcastle upon Tyne NE1 7RU, UK
\aff{2} Laboratoire Atmosph\`eres Milieux Observations Spatiales (LATMOS/IPSL), Sorbonne Universit\'e, UVSQ, CNRS, Paris, FRANCE
\aff{3} {D\'epartement de Math\'ematiques et Applications, UMR-8553, \'Ecole Normale Sup\'erieure, CNRS, PSL University, 75005 Paris, FRANCE}}

\begin{document}

\maketitle


\begin{abstract}
  We consider axisymmetric Boussinesq convection in a shallow cylinder radius, $L$, and depth, $H\,(\ll L)$, which rotates with angular velocity $\Omega$ about its axis of symmetry aligned to the vertical. Constant heat flux boundary conditions, top and bottom, are adopted, for which the onset of instability occurs on a long horizontal length scale provided that $\Omega$ is sufficiently small. We investigate the nonlinear development by well-established two--scale asymptotic expansion methods. Comparisons of the results with the direct numerical simulations (DNS) of the primitive governing equations are good at sufficiently large Prandtl number,~$\sigma$. As $\sigma$ is reduced, the finite amplitude range of applicability of the asymptotics reduces in concert. Though the large meridional convective cell, predicted by the DNS, is approximated adequately by the asymptotics, the azimuthal flow fails almost catastrophically, because of significant angular momentum transport at small~$\sigma$, exacerbated by the cylindrical geometry. To appraise the situation, we propose hybrid methods that build on the meridional streamfunction $\psi$ derived from the asymptotics.  With $\psi$ given, we solve the now linear azimuthal equation of motion for the azimuthal velocity $v$ by DNS. Our ``hybrid'' methods enable us to explain features of the flow at large Rayleigh number, found previously by Oruba, Davidson \&~Dormy (J.~Fluid Mech., vol.~812, 2017, pp.~890--904).
\end{abstract}
 

\section{Introduction\label{introduction}}


\subsection{Background\label{Background}}

The finite amplitude convection in  a horizontal plane layer of Boussinesq fluid, rotating with constant angular velocity $\Omega$ about an axis normal to the plane and driven by an unstable vertical temperature gradient, is a classical problem of continuing interest. Recently, the study has gained a new focus through its possible applicability to the study of tropical cyclones. For that, \cite{ODD17,ODD18} considered axisymmetric convection in a large aspect ratio (penny shaped) cylinder, radius $L$ and depth $H\,(\ll L)$. Motion consists of two parts: (i)~Meridional flow driven by the buoyancy (measured by the Rayleigh number, $\Ra$), which, in turn, stimulates (ii)~Azimuthal (or swirling) motion, through the action of the Coriolis acceleration (measured by the inverse Ekman number, $E^{-1}=H^2\Omega/\nu$; kinematic viscosity $\nu$. The precise form of the convection depends on the nature of the  top and bottom boundary conditions. \cite{ODD17,ODD18} assumed that the bottom boundary is rigid and the top boundary is stress free. They also assumed that the heat flux across the top and bottom boundaries remains constant, as defined by the unperturbed applied vertical temperature gradient. All these characteristics are summarised in figure~2 of \cite{ODD17}. At moderate Rayleigh numbers they found that nonlinear convection consists of one large elongated meridional cell that extends from the symmetry axis to the outer boundary, together with the linked azimuthal flow driven by the Coriolis force. However, as $\Ra$ is increased and motion intensifies, a region of reversed meridional flow appears near the axis \citep[see][figures~3--5]{ODD18}, a feature commonly found in atmospheric vortices, where it is often referred to as an ``eye''. Our objective here is to explore such convection from an asymptotic point of view, based on the small size of the aspect ratio
\be
\label{epsilon}
\epsilon\,\equiv\,H/L\,(\,\ll\,1)\,.
\ee
Our asymptotic method has its limitations. For, though it leads to an understanding of many aspects of the convection, our approach falls short of explaining the strongly nonlinear eye feature for the following reason. A consequence of the long length scale assumption (\ref{epsilon}) is that at leading order the asymptotic solutions of \S\ref{eps-exp}
have separable form ensuring that the axial profiles at all radii are  similar. Such solutions cannot describe eyes with local eddy structure.

A dominant feature of the meridional flow displayed in figures~3--5 of \cite{ODD18} is the large cell, remarked on above, that extends from the symmetry  axis (possibly corrupted by the eye) to nearly the outer boundary. This is a well-known characteristic of non-rotating Rayleigh--B\'enard convection in a plane layer subject to constant heat flux boundary conditions. When that system is unbounded in the horizontal direction, linear solutions may be sought characterised by a horizontal wave number, $k$. For most convection problems the onset of instability occurs at a finite value of $k=k_c$. However, in the case of constant heat flux boundary conditions, onset is characterised by $k_c=0$. The two length scale, $L\gg H$, feature of the convection has been exploited by \cite{CP80,CCP80} to develop a weakly nonlinear theory based on $\epsilon \ll 1$. Demanding that the horizontal length, $L$, be finite, is a prerequisite for any application of the theory to a confined geometry.

The modus operandi for the non-rotating case is described comprehensively by \cite{CP80}. Essentially, 2-D convection is considered relative to $x$ (horizontal) and $z$ (vertical) coordinates. At lowest order in $\epsilon$, the temperature perturbation $\theta$ from the linear (in $z$) conduction state is assumed to be a slowly varying function of $x$ and $t$ alone, independent of $z$; more precisely $\theta = f(X,\tau)$, dependent on the stretched variables $X=\epsilon x$, $\tau=\epsilon^4 t$. Consistency conditions at higher order in the expansion determine the nonlinear amplitude equation
\bme
\label{CP-amp-eq}
\be\se
\partial_\tau f \,= \,G^\prime
\ee
in a conservation law form \citep[see, e.g.,][]{MC00}, where the prime denotes the $X$--derivative. Here,
\be
G\,=\,-A \mu^2 g - Bg^{\prime\prime}+ Cg^3 - Dgg^\prime \hskip 10mm \mbox{in which}\hskip 8mm g=f^\prime,
\ee
\eme
and where $A$--$D$ are non-negative constants \citep[equation~(3.15)]{CP80} and $\mu^2=\epsilon^{-2}(\Ra-\Ra_c)/\Ra_c$ is a measure of the excess Rayleigh Number, $\Ra-\Ra_c$, above the critical value $\Ra_c$ for a horizontally unbounded layer. Similar conservation law equations have been considered in other convective systems \citep[][]{DS81,CY92,Petal04}. Variants of (\ref{CP-amp-eq}), not in conservation law form, have been studied by \cite{S82} and, in higher dimensions (see (\ref{CP-amp-eq-XY}) below), by \cite{C98}.

The symmetries of (\ref{CP-amp-eq}) are important; the most obvious being the invariance under a shift of $X$. Further, the reflection $X\mapsto -X$ admits two symmetries $f\mapsto \pm f$ with $G\mapsto \mp G$, $g\mapsto \mp g$, $g^\prime\mapsto \pm g^\prime$ and so on. For the case $D=0$, we only have odd powers of $f$ and $g$ in (\ref{CP-amp-eq}) and so, without the spatial reflection, we have the additional symmetry $f\mapsto -f$ with $G\mapsto -G$, $g\mapsto -g$. However, when $D\not=0$ this symmetry is lost, because of the quadratic term $- Dgg^\prime$ in (\ref{CP-amp-eq}$b$). On the one hand, the case $D=0$ occurs when the physical system exhibits up/down symmetry. Solutions for that case have been investigated at very large $\Ra$ by \cite{F99} and compared with results from DNS of the full governing equations. On the other hand, $D\not=0$ occurs when that up/down symmetry is broken. The latter is exactly the situation of interest to us, happening because of our asymmetric boundary conditions, stress-free at the top and rigid at the bottom. These various symmetries have consequences for the steady solutions of (\ref{CP-amp-eq}), namely $G=0$, portrayed in figures~4--6 of \cite{CP80}. For their model, $g$ is a measure of $\psi$ (as it is for us), the streamfunction for the flow. So $g\mapsto -g$ implies $\psi \mapsto -\psi$, which, without reversing the sign of $X$, means a reversal of the flow direction.

The solution of the system (\ref{CP-amp-eq}) requires boundary conditions. On assuming spatial periodicity of $f$, $g$, $G$, multiplication of (\ref{CP-amp-eq}$a$) and various integrations by parts determine
\be
\label{CP-energy-eq}
\tfrac12\dR_\tau\bavd{f^2}\,=\,-\,\bavd{gG}\,=\,A \mu^2 \bavd{g^2} - B\bavd{(g^{\prime})^2}- C\bavd{g^4},
\ee
where $\dR_\tau\equiv\dR/\dR \tau$ and $\bavd{\bullet}$ is the spatial average of $\bullet$ over a periodicity length. Fortuitously, the contribution from $D\bavd{g^2g^{\prime}}=\tfrac13 D\bavd{(g^3)^{\prime}}\,(=0)$ vanishes and the remaining form (\ref{CP-energy-eq}) can be employed to show that the bifurcation from the zero to finite amplitude state is necessarily via a supercritical pitchfork.

\cite{D88} extended the \cite{CP80} approach to the case when the  plane layer rotates rapidly about a vertical axis; he employs the Taylor number, $\Ta=E^{-2}$. The work is not totally comprehensive but does point to an amplitude equation (his proposed equ.~(50), similar to (\ref{CP-amp-eq})). However, in his  equ.~(50), he retains a  quadratic term like $Dgg^\prime$ in (\ref{CP-amp-eq}$b$), which we believe vanishes because he limits his study to  boundary conditions with up/down symmetry. These include stress free boundary conditions, often adopted because of the mathematical simplifications that follow \citep[see, e.g., the  related linear study of][]{Tetal02}.

With rotation, motion can no longer lie in an $x$--$z$ plane, as the effect of the Coriolis acceleration is to stimulate motion in the mutually orthogonal third $y$-direction. So though the convection studied by \cite{D88} has components in all three directions, it is said to be 2-D, as it only depends on two coordinates $x$ and $z$. \cite{C98}, however, went further by investigating fully 3-D motion. For that, he introduced the stretched coordinate $Y=\epsilon y$, in an addition to  $T(\,\equiv \tau)$, $X$ of \cite{CP80}, and extended  the form of (\ref{CP-amp-eq}) to an amplitude equation for $f=\phi(X,Y,T)$.

Whereas, \cite{CP80} defined $\epsilon$ as an ad hoc aspect ratio, \cite{C98} perturbs the constant flux boundary condition, $\partial_z\theta=0$, into one of the Robin type, $\partial_z\theta +\alpha \theta =0$, with $\alpha\ll 1$. On making the choice $\epsilon=\alpha^{1/4}$, Cox derives an amplitude equation (his (3.2)), which, when solved subject to periodic boundary conditions, would appear to be reducible to the form
\bme
\label{CP-amp-eq-XY}
\be
\partial_\tT f \,+\,f\,= \,{\bm\nabla}_{\!\!\tH}{\bm \cdot} \Gv_\tH\,, \hskip 15mm  \gv_\tH\,= \,{\bm\nabla}_{\!\!\tH} f\,,
\ee
\eme
where ${\bm\nabla}_{\!\!\tH}\equiv (\partial_\tX,\partial_\tY)$, and $\Gv_\tH$, like $G$ in (\ref{CP-amp-eq}$a$), is a function of $\gv_\tH$ and its space derivatives. The contribution, $+f$, on the left-hand side of (\ref{CP-amp-eq-XY}$a$), originates from the $\alpha \theta$ term in the Robin boundary condition with $\epsilon$ chosen to ensure that, at the onset of instability, the stretched horizontal critical wavenumber $\epsilon^{-1} k_c$ is order unity. For us, this additional ingredient is an embellishment and, with the $+f$ term ignored, (\ref{CP-amp-eq-XY}) achieves conservation law structure.

To investigate the onset of instability, \cite{C98} studied the 2-D extension (\ref{CP-amp-eq-XY}) of (\ref{CP-amp-eq}) to the rotating case $E^{-1}\not =0$. Essentially, for large Ekman number $E$ the coefficient equivalent to $B$ in (\ref{CP-amp-eq}$b$) is positive. On decreasing $E$, that coefficient decreases and vanishes at some $E=E_c$  (say, dependent on the stress boundary conditions adopted). On decreasing $E$ further, $B$ changes sign and becomes negative. Once that happens, the system becomes unstable to short length scale disturbances and the two length scale assumption no longer applies. A similar conclusion was reached in the analytic study of \cite{D88}, albeit in the symmetric case (upper and lower boundaries stress free), who's results were later confirmed numerically by \cite{Cetal15} as illustrated in their figure~1($a$). This consideration places the limit $E>E_c$ on the applicability of the long horizontal length scale approach.

The main thrust of \cite{C98} was the investigation of pattern formation for which his 2-D formulation was essential. He focused attention on the stability of the rhombic lattice (motivated by the \citealt{KL69} instability, but see \citealt{S85} for up/down asymmetry pertinent to us) and square cells. Our objective is 1-D in nature, since it concerns the axisymmetric flows appropriate to cyclones and other related geophysical flows. For our restricted class of flows, it is far simpler to adapt the original \cite{CP80} development to cylindrical geometry, rather than build on either \cite{D88} or \cite{C98}. Specialising Cox's results to that single co-ordinate geometry is unsatisfactory because additional non-trivial work is needed to obtain our amplitude equation from his general form. Unlike \cite{C98}, we are able to obtain, via our appendices~\ref{linear-problem-sol}--\ref{Differentials-averages}, analytic expressions for the coefficients in the amplitude equation. 


\subsection{Objectives and outline\label{Outline}}

Our primary objective is to apply a variant of the amplitude modulation equation (\ref{CP-amp-eq}) to axisymmetric rotating convection in a thin disc, as formulated in \S\ref{CPS2}. However, in the case of rapid rotation $E\ll 1$, it is well known that the onset of convection occurs on a short $E^{1/3}H$ horizontal length scale. So, by necessity, we need to restrict attention to $E> E_c$, which for our problem is $E_c\approx 0.2274$ (see (\ref{F-2-coefs-infinity-more}$a$)). 

A preliminary restructuring of the \S\ref{CPS2} governing equations is undertaken in \S\ref{CPS2-brief} to prepare for the implementation of the \cite{CP80} expansion procedure in \S\ref{eps-exp}. The lowest order terms are considered in \S\ref{vertical-structure-0-brief}, leading to a linear problem for the vertical $z$--structure, whose solution is summarised in appendix~\ref{linear-problem-sol}. The next order problem is formulated in \S\ref{vertical-structure-1-brief}. The consistency condition for its solution, considered in \S\ref{Closure-brief}, leads to a radial amplitude modulation equation $\pdT{f}\,=\,R^{-1} (R\GC_2)^{\prime}$  (\ref{amp-eq-brief}$a$) (cf., (\ref{CP-amp-eq}$a$)), in which $R=\epsilon r$ is the stretched radius. Here $\GC_2$ (\ref{F-2_brief-reduced}$a$) contains coefficients analogous to $A$--$D$ in (\ref{CP-amp-eq}$b$), which are evaluated from analytic results derived in appendices~\ref{G-coefficients} and~\ref{Differentials-averages}. An amplitude equation of similar structure to the Cartesian type (\ref{CP-amp-eq}) was developed by \cite{D88}. Significantly, his Cartesian symmetry $X \mapsto -X$ is lost in our cylindrical geometry, for which there is no corresponding  $R \mapsto -R$ symmetry. Consequences of this lack of symmetry begin to emerge in~\S\ref{amplitude-modulation-problem}, when the thermal energy balance (\ref{average-heat-eq-2-f}) is considered in~\S\ref{TE}. It contains the extra term $\sigma^{-1}2\FC_{\tWW}\bavd{ R^{-1}g^3}$ with no counterpart in the Cartesian version (\ref{CP-energy-eq}). 
  
The weakly nonlinear analysis of \S\ref{ch-var} builds on the linear solution of \S\ref{linear-problem} and brings into sharp focus, in \S\ref{Small-amplitude-expansion-N}, the complications that occur once the basic state bifurcates. In a non-rotating system, two finite amplitude modes emerge through a pitchfork bifurcation distinguished by the direction of motion in the large meridional cell, essentially identified by the sign of the streamfunction $\psi$. Due to the lack of the reflectional symmetry $R \mapsto -R$ with $\psi\mapsto -\psi$, weak nonlinearity affects $\psi$ differently on the two branches, $\psi\gtrless 0$ of the pitchfork. On increasing the rotation rate from zero, the pitchfork tilts and locally changes its character, becoming a transcritical instability  (see \citealt{GH83}), whose implications are discussed at the end of \S\ref{delta-O1}. The subcritical instability, $\psi>0$, corresponds to upwelling on the axis, as found in the full nonlinear DNS of \cite{ODD17,ODD18}. The question of whether or not such solutions, presumably lying on an upper branch of the ``bent'' pitchfork, are accessible via the amplitude modulation equation (\ref{amp-eq-brief}), is addressed by comparison, in \S\ref{Asym-DNS}, of its solutions with the DNS solutions of the complete governing equations. DNS solutions linked to the stable lower supercritical branch are also found, but  we expect that with increasing rotation rate the upper branch solutions are generally realised upon time-stepping from most initial states.

The comparisons of maximum $|\psi|$ on the flow domain for the non-rotating case with only meridional motion, in \S\ref{non-rotating-DNS}, are good up to large $\Ra$. This is surprising because, on increasing $\Ra$, boundary layers form either on the outer $R=1$ or inner $R=0$ boundaries. In this context, a boundary layer is a region, where the horizonal length scale is comparable to or less than the vertical length scale. Solutions of  (\ref{amp-eq-brief}) cannot capture such boundary layer structure, because there the length scale separation, implicit in the  assumption (\ref{epsilon}), does not hold. The solution in the mainstream outside such boundary layers may or may not provide a useful approximation of the DNS of the complete problem. We emphasise this matter in the final paragraph of \S\ref{obl}.

For the rotating case, considered in \S\ref{rotating-DNS}, the asymptotics only gives good agreement with the DNS at moderate $E \ge O(1)$ and Prandtl number $\sigma \ge O(1)$. The limitation on $E$ is anticipated, because, as previously  noted, the long length scale assumption at the instability bifurcation only applies to $E>E_c\approx 0.2274$. On decreasing the value of $\sigma$, we  find in \S\ref{meridional} that the meridional motion fares moderately well. However, that is not the case for the azimuthal velocity $v$ investigated in \S\ref{azimuthal}, for which inertia has such a strong effect that the long length scale assumption is violated with a consequent failure of the asymptotics. As the meridional motion does not seem to be influenced strongly by the azimuthal flow, we undertake hybrid calculations. That is, we substitute $\psi$, as found by the asymptotics, into the azimuthal momentum equation (\ref{CP2.11-12_brief}$a$), which we solve in isolation by DNS to obtain $v$. In \S\ref{lrge-Ra}, we adjust our hybrid approach to test its worth against the large Rayleigh number DNS of \cite{ODD17}. We end with a few concluding remarks in \S\ref{conclusions}.


\section{The rotating frame extension of the \cite{CP80} problem in cylindrical geometry \label{CPS2}}

Relative to cylindrical polar co-ordinates $(r,\, \varphi,\, z)$, we consider axisymmetric  Boussinesq fluid in disc-shaped container radius $L$, depth $H$ with gravity $-g\zvh$, rotating with angular velocity ${\bm \Omega}=\Omega\zvh$. At time $t$, the fluid has velocity $\uv(r,z,t)=(u,\,v,\,w)$, pressure $p$, viscosity $\nu$, thermal diffusivity $\kappa$. Relative to some appropriate reference temperature, the temperature is $-\beta z+\theta(r,z,t)$. Motion is governed by the equations
\bse
\label{CP2.1-3_brief}
\begin{align}
  \sigma^{-1}\DR_t\uv+\,E^{-1}2\vp{\zvh}{\uv}\,=\,&\,-\,\bfnabla p + \Ra\,\theta \zvh+\nabla^2 \uv
  \hskip12mm   \bigl(  \DR_t=\pdt+\sp{\uv}{\bfnabla}\bigr)\,,\\
  \DR_t\theta\,=\,&\,\sp{\uv}{\hat\zv}+\nabla^2\theta \hskip24mm  \bigl(\sp{\bfnabla}{\uv}\,=\,0\bigr)
\end{align}
\ese
in which units used are distance $H$, time ($t$) $H^2/\kappa$, velocity ($\uv$) $\kappa/H$, temperature perturbation ($\theta$) $\beta H$, and where the Rayleigh, Ekman and Prandtl numbers are 
\bme
\label{CP2.4-5_draft}
\be\te
\Ra\,=\,g\alpha\beta H^4/(\kappa\nu)\,, \hskip 10mm E\,=\,\nu/(H^2\Omega)\,, \hskip 10mm \sigma\,=\,\nu/\kappa\,,
\ee
\eme
respectively.

We apply zero perturbation heat flux and zero mass boundary conditions
\bme
\label{CP2.6_brief}
\be
\sp{\nvh}{\bfnabla}\theta\,=\,0\,,  \hskip 15mm  \sp{\nvh}{\uv}\,=\,0
\ee\eme
(outward unit normal $\nvh$) on all boundaries. In view of incompressibility $\sp{\bfnabla}{\uv}=0$ and the boundary condition (\ref{CP2.6_brief}$b$), there is no total vertical mass flux
\be
\label{zero-mass-flux}
\int_0^{1/\epsilon} rw\,\dR r\,=\,0\,,   
\ee
where $\epsilon=H/L$ (\ref{epsilon}). So on integrating the heat conduction equation (\ref{CP2.1-3_brief}$b$) throughout the entire domain $0<r<\epsilon^{-1}$, $0<z<1$, we deduce that
\be
\label{total-heat}
\int_0^1\int_0^{1/\epsilon} r\theta\,\dR r\,\dR z\,=\,\mbox{const.,}
\ee
independent of $t$.

The upper boundary is assumed to be stress free so that
\bse
\label{bc_brief}
\be
\pdz u\,=\,\pdz v\,=\,0  \hskip 10mm  \mbox{at}  \hskip 5mm  z=1\,,
\ee
while the lower and  outer boundaries are assumed to be rigid
\begin{align}
u\,=\,v\,=\,&\,0 &   \mbox{at} &&    z=&\,0\,,\\
\hskip 40mm
v\,=\,w \,=\,&\, 0   & \mbox{at} &&   r=&\,\epsilon^{-1}\,.\hskip 20mm 
\end{align}
\ese
The asymmetric boundary conditions (\ref{bc_brief}$b$,$c$) correspond to the Case C of \cite{CP80}. It is important to note that their non-dimensionalisation, based on the depth $H=2d$ with boundary conditions at $z=\pm 1$, is different from ours. Since we only consider the asymmetric Case C, our non-dimensionalisation, based on boundaries at $z=0$ and $1$, is a more convenient choice for that system. 

We introduce
\bse
\label{psi-varpi}
\begin{align}
(u,\,v,\,w)\,=\,&\,\bigl(-\,r^{-1}\pdz \psi,\,r\omega,\,r^{-1}\pdr\psi\bigr),\\
\vp{\bfnabla}{\uv}\,=\,&\,\bigl(-\,r\pdz \omega,\,-\,r^{-1}\DC\psi,\,r^{-1}\pdr(r^2\omega)\bigr),
\end{align}
where
\be
\DC=r\pdr\bigl(r^{-1}\pdr\bigr)\,+\,\pdzii\,.
\ee
\ese
Then $r$ times the azimuthal component of the momentum equation for $r\omega$, and $-r^{-1}$ times the azimuthal component of the vorticity equation for $-r^{-1}\DC\psi$ determine
\bse
\label{CP2.11-12_brief}
\begin{align}
\sigma^{-1}\DR_t(r^2\omega)-2E^{-1}\pdz \psi\,=\,&\,\DC(r^2\omega)\,,\\
\sigma^{-1}\bigl[\DR_t(r^{-2}\DC\psi)+\pdz(\omega^2)\bigr]+2E^{-1}\pdz \omega =\,&\,\Ra\, r^{-1} \pdr\theta+r^{-2}\DC^2\psi\,,
\end{align}
\ese
respectively, which are to be solved subject to the boundary conditions
\bse
\label{bc_brief-more}
\begin{align}
\psi\,=\,\pdr (r^{-1}\pdr\psi)\,=\,\pdr\omega\,=\,\pdr\theta\,&=\,0& \mbox{at} &&r\,=&\,0\, & (0&<z<1)\,,\\ 
\psi\,=\,\pdr\psi\,=\,\omega\,=\,\pdr\theta\,&=\,0& \mbox{at} &&r\,=&\,\ell\, & (0&<z< 1)\,,\\ 
\psi\,=\,\pdz\psi\,=\,\omega\,=\,\pdz\theta\,&=\,0& \mbox{at} &&z\,=&\,0\, & (0&<r<\ell)\,,\\ 
\psi\,=\,\pdzii\psi\,=\,\pdz\omega\,=\,\pdz\theta\,&=\,0& \mbox{at} &&z\,=&\,1\, & (0&<r<\ell).
\end{align}
\ese
We find it useful to express the heat conduction equation (\ref{CP2.1-3_brief}$b$) in the form
\bme
\label{varphi*}
\be
\DR_t\theta\,=\,r^{-1}\pdr\varphi\,+\,\pdzii\theta\,, \hskip 10mm   \mbox{where}\hskip 10mm   \varphi\,=\,\psi\,+\,r\pdr\theta
\ee
satisfies the boundary conditions
\be\se
\varphi\,=\,0 \hskip 8mm \mbox{at}  \hskip 8mm   r\,=\,0\,\,\mbox{ and }\,\,\ell\,,
\ee
\eme
implied by (\ref{bc_brief-more}$a$,$b$).


\section{Formulation of the small $\epsilon$ problem\label{CPS2-brief}}

Our formulation and development of the small $\epsilon=H/L$ (\ref{epsilon}) case, as explained in the Introduction \S\ref{introduction}, largely follows \cite{CP80} and is essentially a variant of \cite{D88}. We set $r=\epsilon^{-1}R$, $\pdr=\epsilon \pdR$, $\omega=\epsilon^2E^{-1}\varpi$ and write
\bme
\label{rR}
\be
\uv\,=\,\bigl(-\epsilon R^{-1}\pdz\psi,\,\,\epsilon E^{-1}R\varpi,\,\,\epsilon^{2} R^{-1}\pdR\psi\bigr)\,, \hskip 8mm \varphi\,=\,\psi\,+\,R\pdR\theta\,.
\ee
\eme
As the time scale of interest is very long, we set $t=\epsilon^{-4}T$, $\pdt=\epsilon^4 \pdT$ but base the material derivative $\DR_t=\,\epsilon^2\DST$ on the velocity time scale $\epsilon^{-2}$ such that
\bme
\label{tT}
\be
\DST{\,\bullet}\,=\,R^{-1}\JR\bigl(\psi\,,\,\bullet\,\bigr)+\epsilon^2\pdT{\bullet}\,, \hskip 8mm
\JR\bigl(\psi\,,\,\bullet\,\bigr)\,\equiv\,(\pdR\psi)\,\pdz{\,\bullet}\,-\,(\pdz\psi)\,\pdR{\,\bullet}\,.
\ee
\eme
We also set
\be
\label{Ra-exp}
\Ra\,=\,\Ra_c\,+\,\epsilon^2\mu^2\,,
\ee
where $\Ra_c$ is the critical Rayleigh number for the onset of steady convection in the limit~$\epsilon\to 0$.
      
Following our variable changes (\ref{rR})--(\ref{Ra-exp}), the governing equations (\ref{CP2.1-3_brief}) become
\bse
\label{CP2.1-3-expansion}
\begin{align}
\pdzii\theta \,=\,&\,\epsilon^2\NC_\theta\,,\\
\pdzii\varpi + 2R^{-2}\pdz\psi\,=\,&\,\epsilon^2 R^{-1}\NC_\varpi\,,\\
\pdziv\psi-2E^{-2}R^2\pdz\varpi +\Ra_c R\pdR\theta\,=\,&\,\epsilon^2 R\NC_\psi\,,
\end{align}
\ese
in which the terms $O(\epsilon^2)$ and smaller appear on the right-hand side.
They are
\bse
\label{CP2.1-3-expansion-definitions}
\begin{align}
\NC_\theta\,=\,&\,\DST{\theta} -R^{-1}\pdR\varphi\,,\\
\NC_\varpi\,=\,&\,\sigma^{-1}R^{-1}\DST{(R^2\varpi)} -\DeltaB(R\varpi)\,,\\
\NC_\psi\,=\,&\,\sigma^{-1}R\bigl[\DST{(R^{-2}\DC\psi)}+E^{-2}\pdz\bigl(\varpi^2\bigr)\bigr]\nonumber\\
&\,-2\pdzii\bigl[\DeltaB(R^{-1}\psi)\bigr]-\epsilon^2 \DeltaB^{\!2} (R^{-1}\psi)-\mu^2 \pdR\theta\,,
\end{align}
\ese
in which
\bme
\label{relations}
\begin{align}
\DeltaB\, \bullet\,\equiv\,  &  \pdR\bigl[R^{-1}\pdR\bigl(R\,\bullet\,\bigr)\bigr]\,, & \DC\,\bullet\,\equiv\, & \pdzii\bullet\,+\,\epsilon^2R\, \DeltaB\bigl(R^{-1}\,\bullet\,\bigr) \,.
\end{align}
\eme
The esoteric introduction of $\DeltaB$ anticipates the importance of $R^{-1}\psi$ and $R\varpi$, on which it acts in (\ref{CP2.1-3-expansion-definitions}$b$,$c$) (see particularly (\ref{psi-P}$a$) and (\ref{varpi-W}$a$) below).

Since the definite $z$-integral, the $z$-average, and the difference of the boundary values are used repeatedly, we define
\bme
\label{mean-notation_brief}
\be\te
\bint{ \,\bullet\,}{a}{b}\,\equiv\,\tint_a^b\,\bullet\,\,\dR z\,, \hskip 15mm  
\bav{\,\bullet\,}\,\equiv\,\bint{ \,\bullet\,}{0}{1}\,,\hskip 15mm
\bjump{\,\bullet \,}\,\equiv\,\bullet(1)-\bullet(0)\,.
\ee
\eme
An immediate application is to the $z$-average of the heat conduction equation (\ref{CP2.1-3-expansion}$a$). Since the left-hand side average vanishes, $\bav{\pdzii\theta}=\bjump{\pdz \theta}=0$ (use (\ref{CP2.6_brief}$a$)), the remaining right-hand side average must vanish too, leaving $\bav{\NC_\theta}=0$. The evaluation is simplified by the identity $\bav{\JR\bigl(\psi\,,\,\theta\,\bigr)}=\pdR\bav{\psi\pdz\theta}$ (integrate by parts and note that $\psi=0$ on both $z=0$ and~$1$). Accordingly, the $z$-average of (\ref{CP2.1-3-expansion}$a$) together with (\ref{CP2.1-3-expansion-definitions}$a$) and (\ref{tT}) determine the heat conservation law
\bme
\label{mean-theta-eq_brief}
\be
\epsilon^2\pdT\oav{\theta}\,=\,R^{-1}\pdR(R\GC)\,,  \hskip 15mm      R\GC\,=\,\oav{\varphi}-\bav{\psi\pdz\theta}\,.
\ee
\eme
Here, $\GC$ may be interpreted as radial heat flux, which satisfies 
\be
\label{F-bc_brief}
\GC(0,T)\,=\,\GC(1,T)\,=0\,,
\ee
in view of the boundary conditions (\ref{bc_brief-more}$a$,$b$) and (\ref{varphi*}$c$).

On multiplying (\ref{mean-theta-eq_brief}$a$) by $R$, integrating between $R=0$ and $1$ and applying the boundary conditions (\ref{F-bc_brief}), we obtain
\be
\label{total-heat-new}
\dR_{\tTT\,}\oavd{\theta}\,=\,0
\ee
$\bigl(\dR_{\tTT\,}=\dR/\dR T\bigr)$, equivalent to (\ref{total-heat}),
where 
\be
\label{average-volume}
\oavd{\bullet}\,=\,\int_0^1 \oav{\bullet}\,R\,\dR R 
\ee
is a suitably scaled volume integral. Further, on multiplying  (\ref{CP2.1-3-expansion}$a$) by $\theta$, application of (\ref{average-volume}) determines $\bavd{\theta\pdzii\theta}=\epsilon^2\bavd{\theta\NC_\theta}$. Then use of (\ref{CP2.1-3-expansion-definitions}$a$), followed by various integrations by parts, leads to the total thermal energy balance
\be
\label{average-heat-eq}
\tfrac12\epsilon^2\dR_{\tTT\,}\bavd{\theta^2}\,
=\,-\,\epsilon^{-2}\bavd{(\pdz \theta)^2}\,-\,\bavd{R^{-1}\,\varphi\, (\pdR \theta)}\,.
\ee
Since the term $-\epsilon^{-2}\bavd{(\pdz \theta)^2}$ is negative, the only possible thermal energy source is $-\bavd{R^{-1}\,\varphi\, (\pdR \theta)}$, a feature that emphasises the importance of $\varphi$, also present in (\ref{mean-theta-eq_brief}$b$).

We now consider the angular momentum equation (\ref{CP2.1-3-expansion}$b$). On integration once with respect to $z$,  subject to the boundary conditions $\pdz\varpi=\psi=0$ on $z=1$, it yields
\bse
\label{psi-theta-eq}
\be
\pdz\varpi + 2R^{-2}\psi\,=\,-\,\epsilon^2 R^{-1}\bint{\NC_\varpi}{z}{1},
\ee
which on substitution into (\ref{CP2.1-3-expansion}$c$) determines  
\be
\pdziv\psi+4E^{-2}\psi +\Ra R\pdR\theta\,=\,\,\epsilon^2 R \Bigl[\NC_\psi\,-\,2E^{-2}  \bint{\NC_\varpi}{z}{1} \Bigr].
\ee
\ese


\section{The small $\epsilon$ expansion\label{eps-exp}}

In  this section, we develop expansions of the variables, $\YS=[\theta,\,\psi,\,\varphi,\,\varpi,\,\GC\,](R,z,T)$, in the form $\YS=\YS_0+\epsilon^2\YS_2+\cdots\,$. Our objective is the construction of the amplitude modulation equation (\ref{amp-eq-brief}), stated in the next \S\ref{amplitude-modulation-problem}, where its solution is discussed. The development extends \cite{CP80} with some parallels to \cite{D88}. Since the lowest order solution is of separable form expressable as the Hadamard product $\YS_0=\RS(R,T)\circ\ZS(z)$, the following compact differential operator notations
\bme
\label{d-ops}
\begin{align}
\bullet^\prime\,\equiv\, &\,\pdR\!\bullet\,, &  \dot\bullet \,\equiv\,&\,\dR_z\bullet\,,\\
\hskip 12mm \DSplus{\,\bullet}\,\equiv\,&\,R^{-1}\pdR\bigl(R\,\bullet\,\bigr)\,,&
\DSminus{\,\bullet}\,\equiv\,&\,R\pdR\bigl(R^{-1}\,\bullet\,\bigr)\,,\hskip 12mm 
\end{align}
\eme
($\dR_z =\dR/\dR z$) turn out to be useful.


\subsection{The $O(1)$ problem for the vertical $z$-structure \label{vertical-structure-0-brief}}

The lowest order problem is very simply built on the assumption that thermal diffusion in the radial direction is negligible, with (\ref{CP2.1-3-expansion}$a$) approximated by $\pdzii\theta_0=0$. Integration subject to $\pdz\theta_0=0$ at $z=0$ and $1$ determines
\be
\label{CP3.2_brief}
\theta_0\,=\,f(R,T)\,.
\ee

Then neglecting the right-hand side of (\ref{psi-theta-eq}$b$),  we see that
\bme
\label{psi-P}
\be
R^{-1}\psi_0\,=\,\Ra_c g(R,T) P(z)\,,  \hskip 10mm \mbox{where}\hskip 10mm g=f^\prime
\ee
\eme
(notation (\ref{d-ops}$a$)), provided $P(z)$ solves
\bme
\label{P-problem_brief}
\be\se
\LC(P)\,\equiv\,{\ddddot P}\,+\,4E^{-2}P\,=\,-1
\ee
(notation (\ref{d-ops}$b$)) \citep[cf.][equ.~(25)]{D88}. The boundary conditions (\ref{bc_brief-more}$c$,$d$) require
\be\se
P(0)\,=\,P(1)\,=\,{\dot P}(0)\,=\,{\ddot P}(1)\,=\,0\,.
\ee
\eme
We summarise the solution in appendix~\ref{linear-problem-sol}. It lacks the simplicity of Dowling's eqns.~(26),~(27),
applicable to the case of stress free boundaries.

On  neglecting the right-hand side of (\ref{psi-theta-eq}$a$), we obtain
\bme
\label{varpi-W}
\be\se
R\varpi_0\,\equiv\,2 g(R,T)W(z)\,,
\ee
on use of (\ref{CP3.2_brief}) and (\ref{psi-P}), provided that
\be
{\dot W}\,=\,-\,\Ra_c\,P(z)\hskip 10mm \mbox{giving} \hskip 10mm W\,=\,-\,\Ra_c\bint{P}{0}{z}
\ee
\eme
after integration subject to $W(0)=0$ (\ref{W0and1_brief}$a$), implied by $\varpi=0$ at $z=0$.

On further use of (\ref{CP3.2_brief}) and (\ref{psi-P}), the  lowest order approximation of (\ref{rR}$b$) is
\bme
\label{Q}
\be\se
R^{-1}\varphi_0\,=\,R^{-1}\psi_0\,+\,\pdR\theta_0\,=\,-\,g(R,T){\ddot Q}(z)\,,
\ee
where
\be
{\ddot Q}\,=\,-\,\Ra_cP(z)-1\hskip 10mm \mbox{giving} \hskip 10mm {\dot Q}\,=\,W(z)-z
\ee
\eme
after integration and, without loss of generality, the boundary condition choice ${\dot Q}(0)=0$\,. Hence, on neglect of the left-hand side  of (\ref{mean-theta-eq_brief}$a$), integration of its remaining right-hand side  with respect to $R$ implies $R\GC_0$ is a constant. Then, the boundary conditions (\ref{bc_brief-more}$a$,$b$) and (\ref{varphi*}$c$) establish that $\GC_0=0$. In turn, substitution of (\ref{Q}$a$) into (\ref{mean-theta-eq_brief}$b$), recalling that $\pdz\theta_0=0$ implies $\bav{\psi_0\pdz\theta_0}=0$, yields sequentially
\bme
\label{P-Ra}
\be\te
R^{-1}\oav{\varphi_0}\,=\,\oav{\GC_0}\,=\,, \hskip 13mm  \bav{\ddot Q}\,=\,0\,, \hskip 13mm  \Ra_c\oav{P}\,=\,-1
\ee
\eme
on use of (\ref{Q}$a$,$b$). Performing the integral in (\ref{P-Ra}$b$) gives $\bjump{{\dot Q}}=0$, which, having chosen ${\dot Q}(0)=0$, yields ${\dot Q}(1)=0$. So finally (\ref{Q}$c$) implies that $W(1)=1$ and in summary
\bme
\label{W0and1_brief}
\be\te
W(0)\,=\,0\,, \hskip 15mm   W(1)\,=\,1\,, \hskip 15mm  {\dot Q}(0)\,=\,{\dot Q}(1)\,=\,0 \,.
\ee
\eme
Our $P$, $W$, $Q$ notation is adopted to follow the development in eqs.~(3.8), (3.10) of \cite{CP80}.

Finally, we note the useful $z$-average identities
\bme
\label{W0and1-av_brief}
\be
\bav{{\dot W}\,\bullet}\,=\,-\,\bav{W\,\dot{\bullet\:\!}}+\,\bullet(1)\,, \hskip 15mm
\bav{{\ddot Q}\,\bullet}\,=\,-\,\bav{{\dot Q}\,\dot{\bullet\:\!}}
\ee
\eme
which follow from integration by parts and use of the boundary values (\ref{W0and1_brief}). At this early stage the emergence of ${\dot Q}$ in (\ref{Q}$c$), as a derivative, appears contrived because its integral $Q(z)$ is only determined up to an arbitrary constant of integration. Nevertheless, the way our solution method unfolds, $Q$ itself only appears within the $z$-average $\bav{{\ddot Q}Q}$, which on integration by parts takes the unique value $-\bav{{\dot Q}^2}$ (see (\ref{W0and1-av_brief}$b$). Other useful related results are
\be
\label{Q-def}
\left.\begin{array}{ll}- \bav{{\ddot Q}W}\!\!\!& =\,\bav{{\dot Q}{\dot W}}\\[0.2em]
& =\,-\,\Ra_c\bav{P{\dot Q}}\!\! \end{array}\right\}\,=\,\bav{{\dot Q}({\ddot Q}+1)}\,=\,\bav{\dot Q}\,.
\ee


\subsection{The $O(\epsilon^2)$ problem\label{vertical-structure-1-brief}}

Just as for the $O(1)$ problem, we begin our $O(\epsilon^2)$ study with the heat conduction equation (\ref{CP2.1-3-expansion}$a$), for which its right-hand side $\NC_\theta$ (\ref{CP2.1-3-expansion-definitions}$a$) is determined at leading order by two terms $R^{-1}\JR\bigl(\psi_0\,,\,\theta_0\,\bigr)= -\Ra_cg^2{\dot P}$ and $\pdR\varphi_0=-(Rg)^\prime{\ddot Q}$. The ensuing  $\NC_\theta$ may be integrated with respect to $z$ so that the corresponding integral of (\ref{CP2.1-3-expansion}$a$) gives
\bse
\label{theta-2_z} 
\be
\pdz\theta_2\,= -\Ra_cPg^2+{\dot Q}\,\DSplus g\,,
\ee
which, in view of (\ref{P-problem_brief}$b$), (\ref{W0and1_brief}$c$), satisfies the boundary conditions $\pdz\theta_2=0$ at $z=0,\,1$. Multiplication of (\ref{theta-2_z}$a$) by $R^{-1}\psi_0 =\Ra_c Pg$  (\ref{psi-P}$a$) provides the useful result
\be
-\,R^{-1}\psi_0\pdz\theta_2\,=\,\Ra_c \bigl[\Ra_cP^2g^3-P{\dot Q}g\,\DSplus g\bigr]\,.
\ee
\ese
Moreover a further integration of (\ref{theta-2_z}$a$), that notes $-\Ra_c\bint{P}{0}{z}=W$ (\ref{varpi-W}$c$), yields
\bme
\label{theta-2_R}
\be\se
\theta_2\,=\,W g^2\,+\,Q\,\DSplus g\,+\,f_2(R,T)\,,
\ee
where $f_2(R,T)$, like $f(R,T)$ introduced in (\ref{CP3.2_brief}), is at this stage an unknown function, whose value (not needed by us) is only fixed by closure at a higher order. Indeed, since $Q(z)$ is only defined up to an arbitrary constant $\QS$, the corresponding contribution $\QS{\,\DSplus g}$ may be absorbed by $f_2$. The radial derivative of (\ref{theta-2_R}$a$) determines
\be
\pdR\theta_2\,=\,2W g g^\prime\,+\,Q\,\DeltaB g\,+\,g_2\,, \hskip 10mm  g_2\,=\,f^\prime_2\,,
\ee
\eme
where we have recalled that $\pdR\,\DSplus = \DeltaB$ (see (\ref{relations}$a$), (\ref{d-ops}$c$)).

Our next objective is to solve the inhomogeneous equation (\ref{psi-theta-eq}$b$) for $\psi_2$. The leading order terms on its right-hand side are determined from
\bse
\label{NC-varpi-psi}
\begin{align}
\NC_\varpi\,=\,&\,2\sigma^{-1}\Ra_c\,(P{\dot W}-{\dot P}W)\,g\,\DSplus g\,-\,2W\DeltaB g\,,\\
\NC_\psi\,=\,&\,\sigma^{-1}\bigl[\Ra_c^2\,\bigl(P{\dddot {\left.P\right.}}g\,\DSplus g-\,{\dot P}{\ddot P}\, g\,\DSminus g\bigr)
  +\,8E^{-2}\,W{\dot W}\,R^{-1}g^2\bigr] \hskip 10mm \nonumber\\
&\,-\,2\Ra_c\,{\ddot P}\DeltaB g\,-\,\mu^2 g
\end{align}
\ese
(notation (\ref{d-ops}$c$,$d$)). Together with the additional contribution $-\Ra_c R\pdR\theta_2$ (use (\ref{theta-2_R}$b$)) from its left-hand side, (\ref{psi-theta-eq}$b$) determines
\be
\pdziv\psi_2+4E^{-2}\psi_2\,
=\,R\Bigl[\NC_\psi\,-\,\Ra_c\,\pdR\theta_2 \,-\,2E^{-2}  \bint{\NC_\varpi}{z}{1}\Bigr]
\ee
with $\NC_\psi$, $\NC_\varpi$ given by (\ref{NC-varpi-psi}). The equation must be solved subject to $\psi_2=\partial \psi_2/\partial z =0$ at $z=0$ and  $\psi_2=\partial^2 \psi_2/\partial z^2 =0$ at $z=1$. The solution may be expressed in the form
\bme
\label{Psi-2_brief}
\se
\begin{align}
R^{-1}\psi_2\,=\,&\,P\,\bigl(\Ra_c g_2+ \mu^2g\bigr)+P_\tD\,\DeltaB g +P_{\tW}\, g g^{\prime}\nonumber\\
\,&\,+\sigma^{-1}\big[P_\tPP\,g\,\DSminus g+\bigl(P^+_{\tPP} +P^+_{\tWW}\bigr)\,g\,\DSplus g +P_{\tWW}\,R^{-1}g^2\big]\,.
\end{align}
Here the various $P_\bullet(z)$-functions solve
\de
\begin{align}
\LC\bigl(P_\tD\bigr) \,=\,&\,(4/E^2)\,\bint{W}{z}{1}-\,\Ra_c\bigl(2{\ddot P}\,+\,Q\bigr)\,,&  \LC\bigl(P_\tW\bigr) \,=\,&\,-\,2\Ra_c W\,,\\
 \LC\bigl(P_\tPP\bigr) \,=\,&\,-\,\Ra_c^2\,{\dot P}{\ddot P}\,,& \LC\bigl(P^+_\tPP\bigr) \,=\,&\,\Ra_c^2\,P{\dddot {\left.P\right.}}\,,\\
  \LC\bigl(P^+_\tWW\bigr) \,=\,& \,-\,\bigl(4\Ra_c/E^2\bigr)\,\bint{P{\dot W}-{\dot P}W}{z}{1}\,,& 
\LC\bigl(P_\tWW\bigr) \,=\,&\,\bigl(8/E^2\bigr)\,W{\dot W}
\end{align}
\eme
subject to the boundary conditions $P_\bullet(0)={\dot P}_\bullet(0)=P_\bullet(1)={\ddot P}_\bullet(1)=0$ of (\ref{P-problem_brief}$b$). So, on multiplying each of (\ref{Psi-2_brief}$b$-$g$) by $P(z)$, taking the $z$--average, integrating by parts and noting the property  $\LC\bigl(P\bigr)=-1$ (\ref{P-problem_brief}$a$), we obtain the important result
\be
\bav{ P_\bullet}= -\bav{P_\bullet \,\LC (P)}= -\bav{P \LC (P_\bullet)}
\ee
\citep[an extension of the technique employed in appendix~A of][]{CP80}.

Armed with the result (\ref{Psi-2_brief}$a$), we may now use (\ref{psi-theta-eq}$a$) to obtain
\be
R\pdz\varpi_2\,=\, -\, 2R^{-1}\psi_2\,-\,\bint{\NC_\varpi}{z}{1}\,,
\ee
which, upon integration  subject to $\varpi_2=0$ at $z=0$, determines $\varpi_2$. However, that result is not needed to close our problem, as we now demonstrate.


\subsection{Closure\label{Closure-brief}}

The amplitude equation for $f$ follows from (\ref{mean-theta-eq_brief}$a$,$b$), which at lowest order yields
\bse
\label{mean-theta-eq_brief-0}
\begin{align}
  \pdT \oav{\theta_0}\,=\,&\,\DSplus \GC_2\,\equiv\,R^{-1}(R\GC_2)^\prime\,,   \\
R\GC_2\,=\,\oav{\phi_2}\,-\,\bav{\psi_0\pdz\theta_2}\,=\,&\,\oav{\psi_2}\,+\,R\pdR\oav{\theta_2}\,-\,\bav{\psi_0\pdz\theta_2}\,.
\end{align}
\ese
The terms on the right-hand side of (\ref{mean-theta-eq_brief-0}$b$) are determined respectively by the mean values of (\ref{Psi-2_brief}$a$), (\ref{theta-2_R}$b$) and (\ref{theta-2_z}$b$). Collecting them together and noting that the two terms involving $f_2^\prime$ cancel, because $\Ra_c \oav{P}=-1$ (\ref{P-Ra}$c$) implies $(1+\Ra_c \oav{P})f_2^\prime=0$, we are left with
\begin{align}
\GC_2\,=\,&\,-\,\mu^2 \Ra_c^{-1} g - \FC_\tD\,\DeltaB g-\FC^\tW_\tQ g g^\prime-\FC_\tQ\,g\,\DSplus g +\FC_\theta g^3\nonumber\\
\,&\,-\sigma^{-1}\bigl[\FC_{\tPP}\,g\,\DSminus g +\bigl(\FC^+_{\tPP}+\FC^+_{\tWW}\bigr)\,g\,\DSplus g +\FC_{\tWW} R^{-1}g^2\bigr].
\label{F-2_brief}
\end{align}
Here, the coefficients of the terms independent of $\sigma$ are
\bme
\label{F-2-coefs*_brief}
\begin{align}
&\hskip 0mm \Ra_c^{-1}=\,-\,\oav{P}\,,&  \FC_\tD\,=\,&\,-\,\oav{P_\tD}\,-\,\oav{Q}\,,\\
&\left.\begin{array}{l}
\FC^{\tW}_{\tQ}\,=\,-\,\oav{P_{\tW}}-2\oav{W}\,,\\[0.3em]
\FC_{\tQ}\,=\,\Ra_c  \bav{P{\dot Q}}\,=\,-\,\bav{\dot Q}, \end{array}\right\}   &    \FC_\theta\,=\,
&\left\{\begin{array}{l}\Ra_c^2 \bav{P^2}\,,\\[0.3em]
\Ra_c\bav{{\dot P}W}\,,\end{array}\right.
\intertext{where the reductions in (\ref{F-2-coefs*_brief}$c$,$d$) have respectively involved (\ref{Q-def}), (\ref{varpi-W}$b$). The remaining coefficients of the terms proportional to $\sigma^{-1}$ are}
&\hskip 2mm \FC_{\tPP}\,=\,-\,\oav{P_\tPP} \,, & \FC^+_{\tPP}\,=\,&\,-\,\oav{P^+_{\tPP}}\,,\\
&\hskip 1mm \FC^+_{\tWW}\,=\,-\,\oav{P^+_{\tWW}}\,,&  \FC_{\tWW}\,=\,&\,-\,\oav{P_{\tWW}}\,.
\end{align}
\eme
Each of  the six $\bav{ P_\bullet}=-\bav{P \LC (P_\bullet)}$ in (\ref{F-2-coefs*_brief}$b$,$c$,$e$-$h$) are evaluated following various integrations by parts and repeated use of (\ref{varpi-W}$b$,$c$), (\ref{Q}$b$,$c$), (\ref{Q-def}), so giving
\bse
\label{F-2-coefs*-more_brief}
\begin{align}
\FC_\tD\,=\,&\,2\Ra_c\bav{{\dot P}^2}\,-\,\bav{{\dot Q}^2}\, -\,\RC_c^{-1}\bav{W^2}\nonumber\\
\,=\,&\,2\Ra_c\bav{{\dot P}^2}\,-\,(1+\RC_c^{-1})\bav{W^2}\,+\,2\bav{zW}-\tfrac13\,,\\
\tfrac12\FC^{\tW}_{\tQ}\,=\,\FC_{\tQ}\,=\,&\,-\bav{\dot Q}\,=\,-\oav{W}\,+\,\tfrac12\,,\\
\tfrac12 \FC^+_{\tPP}\,=\,\FC_{\tPP}\,=\,&\,\tfrac12 \Ra_c^2\bav{{\dot P}^3}\,,\\
\tfrac23 \FC^+_{\tWW}\,=\,\FC_{\tWW}\,=\,&\,-\,\bigl(4/E^2\bigr)\bav{{\dot P}W^2}\,=\,-\,\bigl(8/E^2\bigr)\Ra_c\bav{P^2W}\,,
\end{align}
\ese
where, in (\ref{F-2-coefs*-more_brief}$a$), we have introduced the alternative measure
\be
\label{RCc}
\RC_c\,=\,E^2\Ra_c\big/ 4
\ee
of $\Ra_c$. Aided by the identities (\ref{F-2-coefs*-more_brief}$b$-$d$), we may reduce (\ref{F-2_brief}) to 
\bse
\label{F-2_brief-reduced}
\be
\GC_2\,=\,-\,\mu^2 \Ra_c^{-1} g - \FC_\tD\,\DeltaB g-\FC_{\tQ\sigma}\,g\bigl(3g^\prime+R^{-1}g\bigr)-\sigma^{-1}2\FC_{\tWW} R^{-1}g^2+\FC_\theta g^3\,,
\ee
where
\be
\FC_{\tQ\sigma}\,=\,\FC_\tQ+\sigma^{-1}\bigl(\FC_{\tPP}+\tfrac12\FC_{\tWW}\bigr)\,.
\ee
\ese

In the non-rotating case $E^{-1}=0$, the $E=\infty$ coefficient $\FC_{\tWW}(E)$ in (\ref{F-2_brief-reduced}) vanishes. The others are linked to $A$--$D$, introduced in (\ref{CP-amp-eq}$b$), and their values follow from Table~1, Case~C of \cite{CP80}, which after appropriate scaling (different units) yield
\bme
\label{F-2-coefs-infinity}
\se
\begin{align}
 \Ra_c(\infty)\,=\,&\,16/A\,=\,320\,,  \\
\FC_\tD(\infty)\,=\,&\,B/4\,=\,58/693\,, \\
\FC_\theta(\infty)\,=\,&\,C\,=\,760/567\,, \\
\FC_{\tQ\sigma}(\infty)\,=\,&\,D/6\,=\,1/18\,+\, (5/126)\sigma^{-1}
\end{align}
composed of
\be\de
\FC_\tQ(\infty)\,=\,1/18\,, \hskip 15mm \FC_{\tPP}(\infty)+\tfrac12\FC_{\tWW}(\infty)\,=\,5/126\,.
\ee
\eme


\begin{figure}
\centerline{}
\vskip 3mm
\centerline{
\includegraphics*[width=1.0 \textwidth]{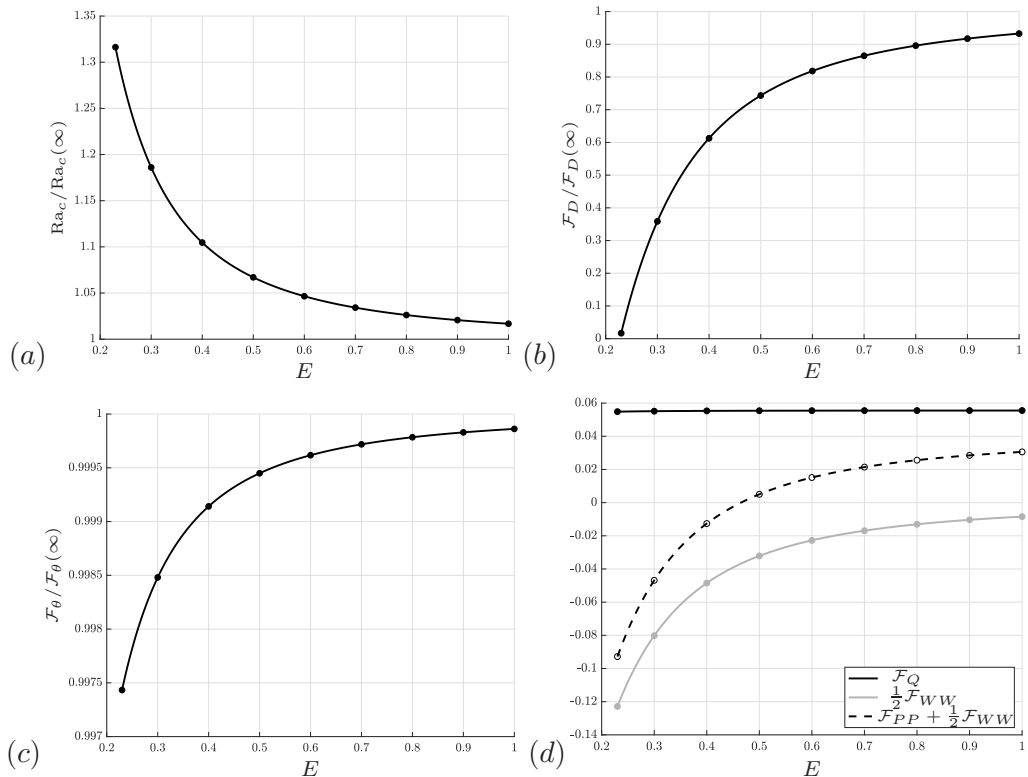}
}
\caption{Plots of (a) $\Ra_c/\Ra_c(\infty)$, (b) $\FC_{\tD}/\FC_{\tD}(\infty)$, (c) $\FC_\theta/\FC_\theta(\infty)$ and (d) $\FC_{\tQ}$ (solid black), $\tfrac12 \FC_{\tWW}$ (solid grey) and $\FC_{\tPP}+\tfrac12 \FC_{\tWW}$ (dashed black) {\itshape versus} $E$ on the range $E>E_c\approx 0.2274$; $\FC_\tD(E_c)=0$ (see (\ref{F-2-coefs-infinity-more}$a$)).}
\label{fig1}
\end{figure}


For the finite $E$ rotating case, the linear problem (\ref{P-problem_brief}) is addressed in appendix~\ref{linear-problem-sol} by considering the Ekman layer style equations (\ref{UV_eq_brief*}) for velocities $\UC(z)$, $\VC(z)$ (\ref{zero-order_sol_brief}), which relate to ${\dot P}$, $W$ (\ref{zero-order_sol-before_brief*}$c$,$b$). Since all the $\FC_\bullet(E)$-coefficients (\ref{F-2-coefs*_brief}$d$) and (\ref{F-2-coefs*-more_brief}$a$-$d$) needed to define $\GC_2$ (\ref{F-2_brief-reduced}$a$) depend on ${\dot P}$ and $W$, we are able to determine their values in terms of  $\UC$ and $\VC$ in appendix~\ref{G-coefficients}. Remarkably, the needed $z$-averages (\ref{PW-UV*}), as well as $\Ra_c^{-1}=4E^{-2}\RC_c^{-1}$ (see (\ref{F-2-coefs*_brief}$a$) and (\ref{RCc})) determined by (\ref{zero-order_sol-more_brief-extra*}), may be expressed entirely in terms of the end point values (\ref{UV_bc_brief*}), (\ref{UV-summary_brief*}) at $z=0$~and~$1$ of the linear solution.
The derivation of the integral results (\ref{linear-means_brief*})--(\ref{UV-cubic-results*}) is relegated to appendix~\ref{Differentials-averages}.

The explicit formulae assembled in appendices~\ref{linear-problem-sol}--\ref{Differentials-averages} show that
\bme
\label{F-2-coefs-infinity-more}
\se
\be
\FC_\tD\gtrless\,0 \hskip 8mm  \mbox{when}\hskip 8mm   E\,\gtrless\,E_c\,\approx 0.2274\,,
\ee
in agreement with the value $(E^{-1}=)\,\Omega_1=4.3966$ given on p.~1347 of \cite{C98}. The positivity 
\be
\FC_\theta\,>\,0     
\ee
is guaranteed by (\ref{F-2-coefs*_brief}$d$). We also have
\be
\FC_{\tWW}\,<\,0\,,
\ee
increasing monotonically through negative values to zero, as $E\to \infty$. Moreover
\be
\FC_\tQ\,>\,0
\ee
increasing from zero at $E=0$ monotonically to $1/18$ (\ref{F-2-coefs-infinity}$e$), as $E\to \infty$, while
\be
\FC_{\tPP}+\tfrac12\FC_{\tWW}\gtrless\,0 \hskip 8mm  \mbox{when}\hskip 8mm  E\,\gtrless\,{\bar E}\,\approx\,0.4650\,,
\ee
\eme
specifically increasing from $-0.5$ at $E=0$ monotonically to $5/126$ (\ref{F-2-coefs-infinity}$f$), as $E\to \infty$. All the behaviours (\ref{F-2-coefs-infinity-more}) pertain to the plots of $\FC_\bullet$ versus $E$ in figure~\ref{fig1}. Each plot is restricted to the  range $E> E_c$, where $\FC_\tD > 0$ (see figure~\ref{fig1}$b$), necessary for the application of our long radial length scale asymptotic assumption. The values of $\Ra_c(E)$, $\FC_\tD(E)$ and $\FC_\theta(E)$ portrayed in figures~\ref{fig1}($a$--$c$) are normalised by their $E\to\infty$ values (\ref{F-2-coefs-infinity}$a$--$c$).

Since the algebra required to determine the results described in appendices~\ref{linear-problem-sol}--\ref{Differentials-averages} is so intricate, we undertook a numerical check for some specific values of $E$. That involved the direct numerical solution of (\ref{P-problem_brief}) for $P(z)$ including, of course, $\Ra_c=-1/\oav{P}$ (\ref{P-Ra}$c$). Whence, the values of the other $\FC_\bullet(E)$-coefficients in (\ref{F-2-coefs*_brief}), (\ref{F-2-coefs*-more_brief}) needed for (\ref{F-2_brief-reduced}) were obtained directly by numerical integration. The results for the selected $E$-values are identified by the bullet points on figure~\ref{fig1}, in perfect agreement with the analytic results.


\section{Amplitude modulation. I.~The problem \label{amplitude-modulation-problem}}

To recap, the heat conservation law  (\ref{mean-theta-eq_brief}$a$) leads to the amplitude equation
\bme
\label{amp-eq-brief}
\be\se
\pdT{f}\,=\,\DSplus\GC_2\,\equiv\,R^{-1} (R\GC_2)^{\prime}
\ee
(\ref{mean-theta-eq_brief-0}$a$), with $\GC_2$ defined by (\ref{F-2_brief-reduced}). It is to be solved subject to some given initial temperature $\theta_0=f(R,0)$ and, for $T>0$, the vanishing heat flux boundary conditions
\be
g(0,T)\,=\,g(1,T)\,=\,0\,, \hskip 10mm \GC_2(0,T)\,=\,\GC_2(1,T)\,=\,0
\ee
\eme
at $R=0$ and $1$. The former (\ref{amp-eq-brief}$b$) identifies zero diffusive flux $\pdR\theta_0=f^\prime=g=0$, which is fortuitously consistent with the kinematic boundary condition $\psi_0=\Ra_c\,R g P(z)=0$. The latter (\ref{amp-eq-brief}$c$) then follows as explained below (\ref{F-bc_brief}).


\subsection{Axial and outer boundary layer considerations\label{obl}}

In addition to the thermal and kinematic boundary conditions (\ref{amp-eq-brief}$b$,$c$), the equations of motion (\ref{CP2.11-12_brief}) are subject to stress boundary conditions  embedded within (\ref{bc_brief-more}). Relevant to that are the tangential components of velocity 
\bme
\label{velocities}
\be
R^{-1}\pdR\psi_0\,=\,\Ra_c\,\DSplus g\, P(z)\,,\hskip 12mm   R\varpi_0\,=\,2g\, W(z)\,, \hskip 10mm
\ee
\eme
and the vertical and azimuthal stresses  proportional to 
\bme
\label{stresses}
\be
\pdR (R^{-1}\pdR\psi_0)\,=\,\Ra_c\DeltaB g\, P(z)\,,\hskip 15mm  R \pdR\varpi_0\,=\,2\,\DSminus g\, W(z)\,,
\ee
\eme
all on a cylinder $R=\,$const. Their appropriate application almost certainly leads to a viscous layer near the outer boundary $R=1$  of radial extent $1-R=O(\epsilon)$, i.e., in a relatively small roughly square region, not accessible by our asymptotics. Though consideration of this layer is needed to determine the solution in the boundary layer, it ought not to influence the ``mainstream'' solution elsewhere at leading order and so we consider it no further.

As our solutions of the amplitude equation (\ref{amp-eq-brief}) have $g\propto R$ as $R\downarrow 0$, the vertical velocity and angular velocity determined  by (\ref{velocities}) are finite on the axis $R=0$, while in turn the stresses (\ref{stresses}) vanish there, as required. The outer boundary $R=1$ is more interesting. Consideration of the expression  (\ref{F-2_brief-reduced}$a$) for $\GC_2$ shows that together $g(1,T)=0$ and $\GC_2(1,T)=0$ (\ref{amp-eq-brief}$b$,$c$) imply
\be
\label{DeltaB-g-0}
\DeltaB g\,=\,0 \hskip 10mm \mbox{at}\hskip 10mm R\,=\,1\,.
\ee
This means that whereas the azimuthal velocity (\ref{velocities}$b$) is brought to rest ($g = 0$), as required by (\ref{bc_brief-more}$b$), the vertical velocity (\ref{velocities}$a$) is not (${\,\DSplus g}\not=0$) contrary to (\ref{bc_brief-more}$b$). Interestingly, a similar problem would arise in the case of a stress free outer boundary. In that case, the vertical stress (\ref{stresses}$a$) vanishes ($\DeltaB g=0$), while the azimuthal stress (\ref{stresses}$b$) does not (${\,\DSminus g}\not=0$, essentially $g^\prime\not=0$ again).

Whether the outer boundary is rigid or stress free, only one (but not both) of the stress boundary conditions can be met and so a boundary layer is required. Interestingly, for the non-rotating problem $E^{-1}=0$, there is no azimuthal flow. So for that case, the problem with a stress free boundary $\DeltaB g=0$ (see (\ref{stresses}$a$) and (\ref{DeltaB-g-0})) at $R=1$ does not require a boundary layer, whereas the case of a rigid boundary, needing ${\,\DSplus g}=0$, does.

We cannot overemphasise our assumption that the $R$-length scale is large compared to $\epsilon$. So whenever solutions of (\ref{amp-eq-brief}) vary significantly on that relatively short $\epsilon$-length scale, i.e., the vertical extent, our asymptotic assumption is violated and the solution of (\ref{amp-eq-brief}) must be viewed with suspicion. The worth of such solutions can only be assessed by  comparison with the DNS of the complete problem, a matter that we address in  \S\ref{Asym-DNS}.


\subsection{The thermal energy balance (\ref{average-heat-eq})\label{TE}}

Our understanding of the nature of the convection and flow is aided by consideration of the thermal energy equation (\ref{average-heat-eq}). The fact that $\theta_0$ (\ref{CP3.2_brief}) is independent of $z$, implying $\pdz \theta_0=0$, has important consequences, which include $\oav{\varphi_0}=0$ (\ref{P-Ra}$a$). In turn, the leading order terms on the right-hand side of (\ref{average-heat-eq}) vanish,
\bme
\label{average-heat-eq-0*}
\be
-\,\epsilon^{-2}\bavd{\theta_0^2}\,-\,2\bavd{\theta_0\theta_1}\,=\,0\,,\hskip 10mm
-\,\bavd{R^{-1}\,\varphi_0 (\pdR \theta_0)}\,=\,0
\ee
\eme
leaving only $O(\epsilon^2)$ terms. What remains, involving  $\varphi_2=\psi_2+R\pdR\theta_2$ (see (\ref{rR}$b$)), is
\be
\label{average-heat-eq-2}
\tfrac12 \dR_{\tTT\,}\bavd{\theta_0^2}\,=\,-\,\bavd{(\pdz \theta_2)^2}\,-\,\bavd{(R^{-1}\varphi_0(\pdR \theta_2)}
\,-\,\bavd{(R^{-1}\psi_2+\pdR\theta_2)(\pdR \theta_0)}\,.
\ee
Aided by the expressions (\ref{theta-2_z}$a$) for $\pdz\theta_2$, (\ref{theta-2_R}$b$) for $\pdR\theta_2$  and (\ref{Psi-2_brief}$a$) for $R^{-1}\psi_2$, the right-hand side may be evaluated tediously. A more direct derivation of the result
\begin{align}
\tfrac12 \dR_{\tTT\,}
\bavd{f^2}\,&\,=\,-\,\bavd{g\GC_2}\nonumber\\
&\,=\,\mu^2 \Ra_c^{-1}\bavd{g^2}\,-\,\FC_\tD\,\bavd{(\DSplus g)^2}\,+\,\sigma^{-1}2\FC_{\tWW}\bavd{ R^{-1}g^3}-\FC_\theta \bavd{g^4}\,, \label{average-heat-eq-2-f}
\end{align}
follows from evaluating the weighted average $\bavd{f\pdT{f}}$ using (\ref{amp-eq-brief}$a$) and integrating by parts. The resulting integral is evaluated using the formula (\ref{F-2_brief-reduced}$a$) for $\GC_2$. In it, the term with the  coefficient $\FC_{\tQ\sigma}$ evaporates because $\bavd{g^2\bigl(3g^{\prime}+R^{-1}g\bigr)}= \bavd{R^{-1}\bigl[Rg^3\bigr]^{\prime}} =0$. Evidently, instability is driven by the term $\mu^2 \Ra_c^{-1}\bavd{g^2}$, when $\mu^2>0$, and damped by the term $-\FC_\theta \bavd{g^4}\,(\,<0)$ (see (\ref{F-2-coefs-infinity-more}$b$)). The diffusive term $-\FC_\tD\bavd{(\DSplus g)^2}$ only damps when $E>E_c$ ($\FC_\tD>0$), otherwise when $E<E_c$ ($\FC_\tD<0$) it drives the instability (see (\ref{F-2-coefs-infinity-more}$a$)). The sign of $\bavd{ R^{-1}g^3}$ in the term $\sigma^{-1}2\FC_{\tWW}\bavd{ R^{-1}g^3}$ is important in determining the nature of the convection, as we argue in the following paragraph. Further consequences are highlighted by our weakly nonlinear theory of \S\ref{Small-amplitude-expansion-N} below.

Typically the meridional flow consists of a single (horizontally elongated) cell, for which the direction of circulation may be identified by the sign of the $z$-average of the scaled vertical velocity, namely 
\be
\label{mean-w*}
\WC(R,T)\,=\,\bav{R^{-1}\pdR \psi_0}\,=\,\bigl(\DSplus g\bigr) \Ra_c\oav{P}\,=\,-\,\DSplus g 
\ee
(use (\ref{P-Ra}$c$)), evaluated on the axis $R=0$. There, (\ref{mean-w*}) determines
\be
\label{up-down}
\WC(0,t)\,=\,-\,\tfrac12 g^\prime(0,T)\,\begin{cases} >\,0\hskip 15mm \mbox{upwelling}\,,\\[0.2em]
<0 \hskip 15mm \mbox{downwelling}\,.\end{cases}
\ee
So, for a single cell with $g(R,T)<0\,(>0)$ on $0<R<1$, we have upwelling (downwelling) on the axis. With that scenario $\bavd{ R^{-1}g^3}<0\,(>0)$,  and since $\FC_{\tWW}<0$ (\ref{F-2-coefs-infinity-more}$c$), the term $\sigma^{-1}2\FC_{\tWW}\bavd{ R^{-1}g^3}>0\,(<0)$ renders the upwelling state to be preferred. This term, however, vanishes in both the infinite Prandtl number limit $\sigma\to 0$ and the non-rotating limit $E\to \infty$ for which $\FC_{\tWW}\uparrow 0$.


\section{Amplitude modulation. II.~The bifurcation,  for the case $E>E_c$\label{ch-var}}

For $E>E_c$,  where $\FC_\tD>0$ (see (\ref{F-2-coefs-infinity-more}$a$)), we reduce the number of independent parameters and highlight the role of various terms by the introduction of the scaled variables (remember too that $\FC_\theta>0$)
\bme
\label{f-nd}
\be\te
T\,=\,\FC_\tD^{-1}\,\tS\,, \hskip 10mm (f,\,g)\,=\,\,-\,\sqrt{\FC_\tD/\FC_\theta}\,\,(\fS,\,\gS)\,, \hskip 10mm \GC_2\,=\,-\,\sqrt{\FC_\tD^3/\FC_\theta}\,\,\GS\,.
\ee
\eme
Since $R^{-1}\psi_0=\Ra_c g(R,T) P(z)$ (\ref{psi-P}$a$) and $\Ra_c\oav{P}=-1$  (\ref{P-Ra}$c$), we expect $\psi_0$ and $g$ to take opposite signs. To avoid that anomaly, we have reversed signs in (\ref{f-nd}$b$,$c$). In terms of the new variables, (\ref{amp-eq-brief}$a$) becomes
\bme
\label{amp-eq-brief-nd}
\be
\partial_{\,\tS\,}{\fS}\,=\,\DSplus\GS\,\equiv\,R^{-1} (R\,\GS)^{\prime} \hskip 10mm \mbox{with}
\hskip 10mm \partial_{\,\tS\,}{\gS}\,=\,\DeltaB\GS
\ee
on differentiation with respect to $R$  ($\fS^\prime=\gS$, see (\ref{psi-P}$b$)), where
\be\se
\GS\,=\,-\,\lambda \gS -\,\DeltaB \gS\,+\,\alpha \gS(3\gS^{\prime}+R^{-1}\gS)\,-\,(\beta/\sigma) R^{-1}\gS^2\,+\,\gS^3\,,
\ee
\eme
in which
\bme
\label{amp-eq-brief-coefs}
\be
\lambda\,=\,\dfrac{\mu^2}{\Ra_c\,\FC_\tD}\,=\, \dfrac{\Ra\,-\,\Ra_c}{\epsilon^2\Ra_c\,\FC_\tD} \hskip 8mm\mbox{equivalently}
\hskip 8mm   \Ra\,=\,\Ra_c\bigl(1+\epsilon^2\lambda\FC_\tD\bigr)
\ee
(see (\ref{Ra-exp})) and
\be
\alpha\,=\,\dfrac{\FC_{\tQ\sigma}}{\sqrt{\FC_\tD\FC_\theta}}\,, \hskip 15mm \beta\,=\,-\,\dfrac{2\FC_{\tWW}}{\sqrt{\FC_\tD\FC_\theta}}\,\,\,(>0)\,.
\ee
\eme

The value of $\alpha$ takes the sign of $\FC_{\tQ\sigma}=\FC_\tQ+\sigma^{-1}\big(\FC_{\tPP}+\tfrac12\FC_{\tWW}\big)$ (\ref{F-2_brief-reduced}$b$). Since $\FC_\tQ\,>0$ (\ref{F-2-coefs-infinity-more}$d$), it follows that  $\FC_{\tQ\sigma}>\sigma^{-1}\big(\FC_{\tPP}+\tfrac12\FC_{\tWW}\big)>0$, when $E>{\bar E}$ (see (\ref{F-2-coefs-infinity-more}$e$)). Of course, $\FC_\tQ$ may exceed zero for smaller $E$, but all our comparisons of asymptotic results with DNS in \S\ref{Asym-DNS} are undertaken  for $E>{\bar E}\,(>E_c)$ and correspond to $\alpha>0$. The positivity of $\beta$ follows because $\FC_{\tWW}<0$ (\ref{F-2-coefs-infinity-more}$c$).  Significantly, $-\FC_{\tWW}\downarrow 0$ implying $\beta \to 0$ as $E\to \infty$, and so it follows that $\beta/\sigma\downarrow 0$ when either $E\to \infty$  or $\sigma\to \infty$.


\subsection{The linear problem\label{linear-problem}}

The linearised version of the amplitude equations (\ref{amp-eq-brief-nd}$a$,$b$) are  
\be
\label{amp-eq-brief-lin}
\partial_{\,\tS\,}{\fS}\,=\,-\,\DSplus(\lambda \gS + \DeltaB \gS )\,, \hskip 15mm \partial_{\,\tS\,}{\gS}\,=\,-\,\DeltaB(\lambda \gS +\DeltaB \gS)\,.
\ee
Note that $\fS$ is only determined up to an arbitrary constant, which we ignore below in order to reduce clutter. The solutions that satisfy the boundary conditions (\ref{amp-eq-brief}$b$,$c$) are
\bse
\label{linear-sol}
\begin{align}
-\,j_m^{-2}\,\DSplus \gS\,=\,\fS\,=\,&\,-\,j_m^{-1}\,A_m(\tS)\JR_0(j_m R)\,,\\
\gS =\,\fS^\prime =\,&\,A_m(\tS)\JR_1(j_m R)\,,\\
\DeltaB \gS\,=\,&\,-\,j_m^2\gS\,,\\
\GS\,=\,-\,\bigl(\lambda \gS +\DeltaB \gS \bigr)\,=\,&\,\bigl(-\lambda+j_m^2\bigr)\gS \,,
\end{align}
\ese
where $j_m$ is the $m^{\tR\hR}$ zero of the Bessel function $\JR_1$, chosen such that $\gS(0,\tS)=\gS(1,\tS)=0$ (see (\ref{linear-sol}$b$)) with the consequence $\GS(0,\tS)=\GS(1,\tS)=0$ by (\ref{linear-sol}$d$), provided that
\be
\label{A-eq}
\dR_{\;\!\tS}{A_m}\,=\,j_m^2 \bigl[\lambda\,-\,j_m^2\, \bigr]A_m  
\ee
$(\dR_{\;\!\tS}\equiv\dR/\dR \tS)$. The requirement $\lambda= \mu^2/(\Ra_c\FC_\tD) > 0$ (see (\ref{amp-eq-brief-coefs}$a$)) for instability is only met when $\FC_\tD>0$, which requires $E>E_c\approx 0.2274$ (see (\ref{F-2-coefs-infinity-more}$a$) and figure~\ref{fig1}($b$)).

The steady modes, $\dR_{\;\!\tS}A_m=0$, correspond to
\be
\label{steady-A}
\lambda\,=\,\lambda_m\,=\,j_m^2\,.
\ee
The lowest mode $m=1$ is identified by the first nonzero zero of $\JR_1$, namely
\bme
\label{lowest-mode}
\be
j_1\,\approx\,3.83171\,, \hskip 12mm  \lambda_1\,\approx\,14.68197\,.
\ee
\eme
Thus, correct to $O(\epsilon^2)$, the critical Rayleigh number determined by (\ref{amp-eq-brief-coefs}$b$) is
\be
\label{Ra-dag}
\Ra_c^\dag
\,=\,\Ra_c(1\,+\,\epsilon^2 j_1^2 \FC_\tD)\,.
\ee

We note that $\psi \propto R\gS \propto R\JR_1(j_1 R)$ is  maximised when $\dR_R [R\JR_1(j_1 R)]=0$, equivalently $\JR_0(j_1 R)=0$. Thus, the first zero of $\JR_0$ determines the location 
\bse
\label{R-max}
\be
R_{\tM 0}\,\approx\,0.6276\,,
\ee
of the maximum, which itself is proportional to $ R_{\tM 0}\JR_1(j_1 R_{\tM 0})$, where
\be
\JR_1(j_1 R_{\tM 0})\,\approx\,0.5191\,.
\ee
\ese
As a corollary, $\fS\propto \JR_1(j_0 R)$ reverses sign across $R_{\tM 0}$.


\subsection{Small amplitude expansion about critical\label{Small-amplitude-expansion-N}}

For our finite amplitude solutions, a useful measure of supercriticality, relative to $\Ra_c^\dag$ (\ref{Ra-dag}), is
\be
\label{aleph}
\aleph\,=\,\dfrac{\lambda-j_1^2}{j_1^2}\,=\,\dfrac{\Ra\,-\,\Ra_c^\dag}{\Ra_c^\dag\,-\,\Ra_c}\,.
\ee

In the following two subsections, we consider a small amplitude expansion
\bme
\label{delta-exp}
\se
\begin{align}
 j_1^2\aleph\,=\,\lambda\,-\,j_1^2\,=\,&\,\delta \Lambda_1 \,+\,\delta^2 \Lambda_2\,+\cdots \,\,,\\
 [\fS,\,\gS]\,-\,\delta[\fS_0,\,\gS_0](R,t)=\,&\,\delta^2 [\fS_1,\,\gS_1](R,t)\,+\,\delta^3 [\fS_2,\,\gS_2](R,t)\,+\cdots
\end{align}
($\epsilon^2\ll \delta\ll 1$) for the lowest steady $m=1$ mode (\ref{linear-sol}$a$,$b$), which solves
\be\de
\square\,\gS_0\,=\,0\,,  \hskip 10mm \mbox{where} \hskip 10mm \square\,\bullet \,\equiv \,\bigl(\DeltaB + j_1^2\,\bigr)\bullet
\ee
\eme
(see (\ref{linear-sol}$c$,$d$)). The objective is to construct the equation governing the slow evolution of the amplitude $A_1(t)$. The positive parameter $\delta\,(\ll 1)$ is chosen at our convenience to  aid identification of the terms, which balance at various orders of $\delta\,(>0)$.


\subsubsection{The case $\beta/\sigma=O(1)$\label{delta-O1}}

For this generic case, we consider only the leading order terms $\delta \Lambda_1$ and $\delta^2 [\fS_1,\,\gS_1]$ on the right-hand sides of (\ref{delta-exp}$a$,$b$). Anticipating evolution on the slow time scale $\delta^{-1}$, we write
\be
\label{delta-O1-exp}
A_1(\tS)=\AS(\TS_{\!1})\,, \hskip 10mm \TS_{\!1}=\delta \tS\,,\hskip 10mm \partial_{\,\tS}=\delta\,\partial_{\,\TS_{\!1}}\,,\hskip 10mm
\dR_{\;\!\tS}=\delta\,\dR_{\TS_{\!1}}\,.
\ee
Then at $O(\delta)$, (\ref{amp-eq-brief-nd}$a$,$c$) determine
\be
\label{delta-O1-exp-con}
R^{-1}\bigl[R \bsquare\gS_1 \bigr]^{\prime}\,=\,-\,\partial_{\,\TS_{\!1}}\fS_0\,+\,R^{-1}\bigl[-\Lambda_1 R\gS_0 +\,\alpha R \gS_0(3\gS_0^{\prime}+R^{-1}\gS_0)    \,-\,(\beta/\sigma)\gS_0^2\bigr]^{\prime}\,,
\ee
(notation (\ref{delta-exp}$d$)), where significantly the cubic term $+\gS^3$ in (\ref{amp-eq-brief-nd}$c$), being smaller by another factor  $O(\delta)$,  has been omitted.

We take the radial weighted average $\bavd{(\ref{delta-O1-exp-con}a)\JR_0(j_1 R)}$ to eliminate the left-hand side and so obtain
\be
\label{delta-O1-exp-eq}
j_1^{-2}\,\dR_{\TS_1\,}\AS \,=\,\Lambda_1\AS \,+\,N_2\,(\beta/\sigma)\AS^2\,.  
\ee
Here, we have used the properties
\bse
\label{delta-O1-exp-prop}
\be
\bavd{\gS^2_0(3\gS_0^{\prime}+R^{-1}\gS_0)}\,=\,\bavd{R^{-1}\dR_R(R\gS_0^3)}\,=\,0\,,
\ee
as  $\gS_0(1,t)=0$, to eliminate the term proportional to $\alpha$, and noted that
\be
\bavd{\JR^2_0(j_1 R)}\,=\,\bavd{\JR^2_1(j_1 R)}\,=\,\tfrac12\JR^2_0(j_1)\,\approx\,0.0811\,.
\ee
In addition, since $\bavd{R^{-1}\JR^3_1(j_1 R)}\approx 0.0821$, we have
\be
N_2\,=\,{\bavd{R^{-1}\JR^3_1(j_1 R)}}\Big/{\bavd{\JR^2_1(j_1 R)}}\, \approx\,1.0124\,.
\ee
\ese

The bifurcation of the steady trivial solutions $\AS=0$ of (\ref{delta-O1-exp-eq}) at $\Lambda_1=0$ to the neighbouring finite amplitude solutions
\be
\label{st-line}
\AS\,=\,-\,\Lambda_1\sigma/(\beta N_2)\,\gtrless \,0\hskip 10mm \mbox{for}\hskip 10mm  \Lambda_1\,\lessgtr\, 0\,,
\ee
since $N_2\,(\beta/\sigma)>0$, is transcritical (see \citealt{GH83} and \citealt{CH93} figure~6). Obviously the solutions, $\AS>0$ (upwelling on the axis) for $\Lambda_1<0$, are unstable and will evolve to a large amplitude for which the weakly nonlinear theory developed here no longer applies. We expand on this matter next.


\subsubsection{The case $\beta/\sigma=O(\delta)$\label{delta-O2}}

To capture the stabilising term  $+\gS^3$ of (\ref{amp-eq-brief-nd}$c$) omitted in (\ref{delta-O1-exp-con}), we consider the case $\beta/\sigma=O(\delta)$. In practice, this limit restricts our analysis to the case $E\gg 1$ of slow rotation, but nevertheless reveals, in more detail, the nature of possible finite amplitude solutions of (\ref{amp-eq-brief-nd}$a$). Since the magnitude of the term $N_2\,(\beta/\sigma)\AS^2$ in (\ref{delta-O1-exp-eq}) is reduced by a factor $O(\delta)$, we reduce  $\aleph$ (\ref{delta-exp}$a$) by the same amount. Accordingly,  we set
\be
\label{aleph-small}
\Lambda_1\,=\,0\,,
\ee
while lengthening the time scale
\be
\label{delta-O2-exp}
A_1(\tS)=\AS(\TS_{\!2})\,, \hskip 10mm \TS_{\!2}=\delta^2 \tS\,,\hskip 10mm \partial_{\,\tS}=\delta^2\,\partial_{\,\TS_{\!2}}\,,\hskip 10mm
\dR_{\;\!\tS}=\delta^2\,\dR_{\TS_{\!2}}\,.
\ee
Then at $O(\delta)$, (\ref{delta-O1-exp-con}) simplifies leaving us with only the $\alpha$-term on the right-hand side. After integration of what remains and application of the end point conditions $\GS=0$ at $R=0$ and $1$, where $\gS_0=0$ too, we obtain
\bse
\label{delta-O2-exp-prob-1}
\be
\bsquare\gS_1 \,=\,\alpha \gS_0(3\gS_0^{\prime}+R^{-1}\gS_0)
\ee
with solution
\be
\gS_1 \,=\,\alpha \gS_0 \fS_0\,=\,-\,\AS^2\alpha j_1^{-1}\JR_0(j_1 R)\JR_1(j_1 R)
\ee
\ese
vanishing at $R=0$ and $1$. In this way, correct to the lowest two orders, we have
\bse
\label{g-delta-sol-sum*}
\begin{align}
\fS\,= \,&\,\delta\fS_0\bigl[1\,+\,\tfrac12\alpha\delta\fS_0\bigr] \,,\\
\gS\,=\,\fS^\prime\,= \,&\,\delta\gS_0\bigl[1\,+\,\alpha\delta\fS_0\bigr].
\end{align}
\ese

Consideration of the maximum of $R(\gS_0+\delta\gS_1)$ reveals a shift in the linear value $R_{\tM 0}$ (\ref{R-max}$a$) for the maximum of $\psi$ to $R_\tM$ given by the solution of
\bse
\label{R-M-dag}
\be
j_1 \JR_0(j_1 R)\,=\,\alpha \delta \AS\bigl[\JR_0^2(j_1 R)-\JR_1^2(j_1 R)\bigr].
\ee
The Taylor series expansion of $\JR_0(j_1 R)$ about $R_{\tM 0}$, at which $\JR_0(j_1 R_{\tM 0})=0$, reveals the lowest order result
\be
R_\tM\,-\,R_{\tM 0}\,=\,\alpha j_1^{-2}  \delta \gS_0(R_{\tM 0})  \,=\,\alpha \delta \AS j_1^{-2}\JR_1(j_1 R_{\tM 0})\,,
\ee
\ese
with $\JR_1(j_1 R_{\tM 0})\approx 0.5191$ (\ref{R-max}$b$). The result (\ref{R-M-dag}$b$)
quantifies the out(in)ward shift of the maximum of $|\psi|$ for solutions that up(down)well, $\AS\,>(<)\,0$, on the axis.

The  $O(\delta^2)$ terms in  (\ref{amp-eq-brief-nd}$a$,$c$) give
\bse
\label{delta-O2-exp-prob-2}
\begin{align}
& R^{-1}\bigl[R \bsquare\gS_2 \bigr]^{\prime}\,=\,-\,\partial_{\,\TS_{\!2}}\fS_0\nonumber\\
&\hskip 18mm +\, R^{-1}\bigl[-\Lambda_2 R\gS_0\,
    +\,\alpha\big(3R(\gS_0 \gS_1)^{\prime}+2\gS_0\gS_1\bigr)\,-\,\delta^{-1}(\beta/\sigma)\gS_0^2
   \,-\,R\gS_0^3\bigr]^{\prime},
\end{align}
cf.~(\ref{delta-O1-exp-con}). As in \S\ref{delta-O1}, we take its radial weighted average $\bavd{(\mbox{\ref{delta-O2-exp-prob-2}})\JR_0(j_1 R)}$ to eliminate the left-hand side. Recalling that $\gS_1=\alpha \gS_0 \fS_0$ (\ref{delta-O2-exp-prob-1}$b$), evaluation of the term proportional to $\alpha$ is aided by the identity
\be
\gS_0   \big(3(\gS_0 \gS_1)^{\prime}+2 R^{-1}\gS_0\gS_1\bigr)\,=\,2\,\DSplus(\gS^2_0\gS_1)\,-\,\alpha\gS^4_0\,.
\ee
\ese
In this way, we obtain
\bme
\label{delta-O2-exp-eq}
\be
j_1^{-2}\,\dR_{\TS_2\,}\AS \,=\,\Lambda_2\AS \,+\,N_2\,(\beta/\sigma)\delta^{-1}\AS^2\,-\,\Upsilon N_3\AS^3\,,
\hskip 12mm \Upsilon\,=\,1+\alpha^2\,,
\ee
\eme
in which 
\be
\label{delta-O2-exp-prop}
N_3\,=\,{\bavd{\JR^4_1(j_1 R)}}\Big/{\bavd{\JR^2_1(j_1 R)}}\,\approx\,0.2517\,,
\ee
since $\bavd{\JR^4_1(j_1 R)}\approx 0.02041$.

The equation (\ref{delta-O2-exp-eq}$a$), albeit only valid when $\beta/\sigma=O(\delta)$, reveals the nature of the bifurcation beyond the transcritical regime identified in \S\ref{delta-O1} for $\beta/\sigma=O(1)$. As the steady finite amplitude solutions $\AS$ satisfy
\be
\label{parabola}
\Lambda_2\,+\,N_2(\beta/\sigma)\AS\,-\,\Upsilon N_3\AS^2\,=\,0\,,
\ee
the transcritical bifurcation at $\Lambda_2=0$, described by (\ref{st-line}), becomes the tangent to the parabola (\ref{parabola}). For that, there are two positive $\AS$ solutions on $\Lambda_{\rm{min}}<\Lambda_2<0$, which coalesce at $\Lambda_2=\Lambda_{\rm{min}}$ with value $\AS=\AS_{\rm{min}}$, where
\bme
\label{A-min}
\be
\Lambda_{\rm{min}}\,=\,-\,\tfrac12 N_2(\beta/\sigma)\AS_{\rm{min}}\,, \hskip 10mm
\AS_{\rm{min}}\,=\,\dfrac{N_2(\beta/\sigma)}{4\Upsilon N_3} \hskip 8mm \bigl(\, N_2(\beta/\sigma)>0\,\bigr).
\ee
\eme
Presumably, for $\Lambda_{\rm{min}}<\Lambda_2<0$, only the upper branch $\AS>\AS_{\rm{min}}$  is stable, whereas, for $\Lambda_2>0$, both the positive and negative  $\AS$ branches are stable. This presumption suggests that large amplitude solutions  of the DNS for the full problem (\ref{CP2.11-12_brief}), (\ref{bc_brief-more}) exist in the generic case $\beta/\sigma=O(1)$, but are not accessible by the weakly nonlinear theory of \S\ref{delta-O1}.


\subsubsection{The non-rotating case $E^{-1}=0$\label{E-1=0}}

Interestingly, the trancritical instability identified by (\ref{st-line}) degenerates when $\beta/\sigma=0$. That happens in the non-rotating case $E^{-1}=0$, upon which we briefly comment here. It is a special case of the previous \S\ref{delta-O2} with the quadratic term $N_2\,(\beta/\sigma)\delta^{-1}\AS^2$ absent from (\ref{delta-O2-exp-eq}$a$). As a consequence, the bifurcation at $\Lambda_2=0$ is a pure pitchfork. For $\Lambda_2>0$, the two steady state solutions $\pm |\AS|$ are determined by the vanishing of what remains of (\ref{delta-O2-exp-eq}$a$), which, noting $j_1^2\aleph=\delta^2\Lambda_2$ (see (\ref{delta-exp}$a$)), gives
\be
\label{amp-delta}
|\delta \AS|^2\,=\,j_1^2\aleph\big/\bigl(\Upsilon N_3\bigr)\,.
\ee

We emphasise that the symmetry of the bifurcation (\ref{amp-delta}), possessing solutions $\pm |\AS|$, is a low order result. Taken to next order, the solution (\ref{g-delta-sol-sum*}), in which $\fS_0$ and $\gS_0$ are proportional to $\AS$, is clearly not invariant under the sign change $\AS\mapsto -\AS$. Moreover,  both correction terms $\alpha\delta\fS_0$ in (\ref{g-delta-sol-sum*}$a$,$b$) change sign, as $R$ crosses $R_{\tM 0}$, at which $\fS_0=0$, as noted below (\ref{R-max}$b$). 
The shift of the maximum (\ref{R-M-dag}$b$) for each, obtained using (\ref{delta-O2-exp-eq}$b$) and (\ref{amp-delta}), is
\be
\label{R-M-dag-E0}
R_\tM\,-\,R_{\tM 0}\,=\,\pm\,\dfrac{1}{j_1}\sqrt{\dfrac{\alpha^2 \aleph}{(1+\alpha^2)N_3}}\,\JR_1(j_1 R_{\tM 0})\,.
\ee


\section{The steady solutions: Asymptotics ($\epsilon\ll 1$) versus DNS\label{Asym-DNS}}

The steady solutions of the reduced asymptotic equation (\ref{amp-eq-brief}$a$), meeting the end point conditions (\ref{amp-eq-brief}$c$), satisfy $\GC_2=0$. In the rescaled units (\ref{f-nd}), the nonlinear problem becomes: Solve $\GS=0$ (see (\ref{amp-eq-brief-nd}$c$)) subject to $\gS=0$ at $R=0$ and $1$ (\ref{amp-eq-brief}$b$). From these we deduce the streamfunction $\psi$ and azimuthal velocity $v$, which we compare with the steady DNS solution. The DNS solution is obtained by time integrating the complete problem (\ref{CP2.11-12_brief}), (\ref{bc_brief-more}) discretised with finite differences until a steady state is reached. We compare the reduced asymptotic and the DNS solution for the case $\ell=10$, i.e.,
\be
\label{eps-0.1}
\epsilon=0.1\,,
\ee
in the following subsections. There, all results displayed in the figures pertain to the unscaled variables $r(=10 R)$, $\psi$ and $v=r\omega$, as they appear in (\ref{psi-varpi}$a$). For each DNS displayed we give $\Ra$, $E$ and $\sigma$.

To formulate the asymptotic amplitude equation $\GS=0$, we need the coefficients  $\lambda$, $\alpha$ and $\beta/\sigma$ appearing in $\GS$ (\ref{amp-eq-brief-nd}$c$). The formula (\ref{amp-eq-brief-coefs}$a$) determines $\lambda$ as a function of $\Ra$  and $E$, while $\alpha$, $\beta/\sigma$ (\ref{amp-eq-brief-coefs}$c$,$d$) are functions of $E$ and $\sigma$. Rather than  $\lambda$, supercriticality may be measured by $\aleph=j_1^{-2}\lambda -1$ (\ref{aleph}).

As announced in the paragraph following (\ref{amp-eq-brief-coefs}) before the start of \S\ref{linear-problem}, all our results pertain to $E>{\bar E}\,(>E_c)$, for which the parameters $\alpha$, $\beta$ are both positive.


\subsection{The non-rotating case, $E^{-1}=0$\label{non-rotating-DNS}}

\begin{figure}
\centerline{}
\vskip 3mm
\centerline{
\includegraphics*[width=1.0 \textwidth]{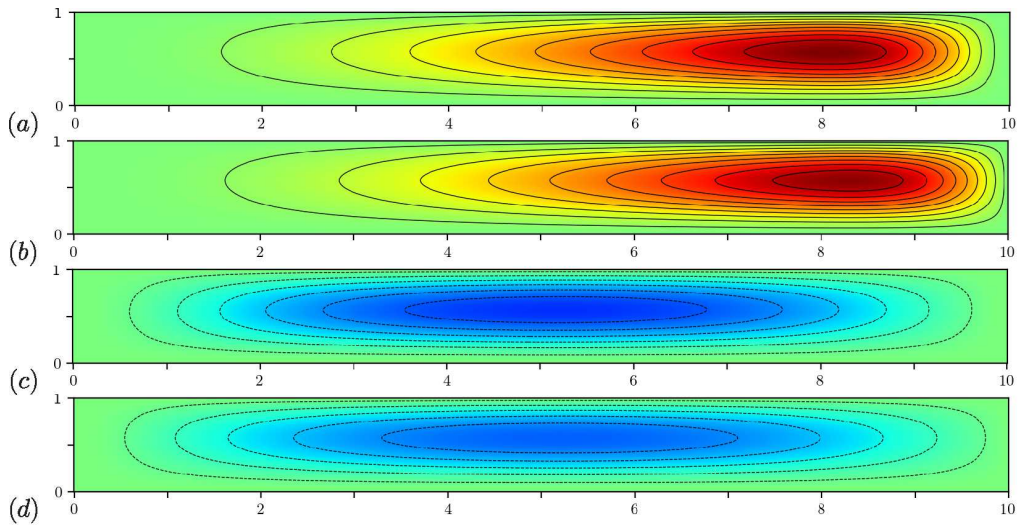}
}
\caption{No rotation case ($E^{-1}=0$), $\sigma=0.1$, $\Ra=330$; a comparison in the $r$--$z$ plane of results obtained from DNS with asymptotics (labelled $A$).\quad   ($a$) $\psi_{\tDNS}>0$; ($b$) $\psi_{\tA}>0$; ($c$)~$\psi_{\tDNS}<0$; ($d$) $\psi_{\tA}<0$  (colour scale from $-1.3$ (blue) through 0 (green) to $1.3$ (red)).}
\label{fig2}
\end{figure}

For the case $E^{-1}=0$, $\sigma=0.1$, $\Ra=330$, we compare in figure~\ref{fig2} the streamlines obtained from the DNS (figures~\ref{fig2}($a$,$c$)) and the asymptotics (figures~\ref{fig2}($b$,$d$)). Relative to the onset values, $\Ra_c=320$, $\Ra_c^\dag\approx 323.9321$, the  supercriticality (\ref{aleph}) of the finite amplitude solution is 
\bme
\label{EO-onset}
\be
\aleph\,\approx\,1.54315  \hskip 10mm  \mbox{equivalently}  \hskip 10mm  \lambda \approx 37.33844\,.
\ee
The remaining $\GS$-coefficients in (\ref{amp-eq-brief-nd}$c$) are $\beta=0$ and
\be\se
\alpha\,\approx\, 0.1659 + 0.11848\sigma^{-1}\,\approx\,1.3506 \hskip5mm  \mbox{for} \hskip3mm \sigma=0.1\,.
\ee
\eme

As we stressed in \S\ref{E-1=0}, the pitchfork bifurcation at $\Ra_c^\dag$ sheds two solutions: one characterised by upwelling on the axis, near which $\psi>0$ (figures~\ref{fig2}($a$,$b$)); the other characterised by downwelling on the axis, near which $\psi<0$ (figures~\ref{fig2}($c$,$d$)). Though the bifurcation is symmetric with the infinitesimal maximum of $|\psi|$ at $R=R_{\tM 0}$, on increasing $\lambda$ that maximum shifts outwards for the $\psi>0$ solutions and inwards for the $\psi<0$ solutions. Such behaviour was predicted by (\ref{R-M-dag}$b$) for the case of small but finite amplitude motion. However, the value $|R_\tM\,-\,R_{\tM 0}|\approx 0.2696$ determined from (\ref{R-M-dag-E0}), though qualitatively plausible, overestimates the shifts visible on figure~\ref{fig2}, because $\aleph \approx 1.54315$ (\ref{EO-onset}$a$) is too large for the small-$\delta$ asymptotics of \S\ref{Small-amplitude-expansion-N} to provide quantitative accuracy. By contrast, the excellent agreement of the DNS (figures~\ref{fig2}($a$,$c$)) with the numerical solutions (figures~\ref{fig2}($b$,$d$)) of
\be
\label{amp-eq-brief-nd-extra}
\bigl(R^{-1}(R\gS)^{\prime}\bigr)^{\prime}\,=\,-\,\lambda \gS \,+\,\alpha \gS(3\gS^{\prime}+R^{-1}\gS)\,+\,\gS^3\,,
\ee
namely $\GS=0$ (see (\ref{amp-eq-brief-nd}$c$)) with $\beta=0$, is most encouraging.

We can make an interesting comparison of the contour plots in figure~\ref{fig2} with those in figure~4($b$) of \cite{CCP80} for their internally heated case exhibiting up/down asymmetry like us. We may capture the structure of the steady state version of their Cartesian asymptotic equation~(15) by dropping the curvature terms in (\ref{amp-eq-brief-nd-extra}) and, where, in our \S\ref{introduction} notation, $R$ has become~$X$. This leaves $\gS^{\prime\prime}=-\lambda \gS +3\alpha \gS\gS^{\prime}+\gS^3$, but note the sign reversal in (\ref{f-nd}$b$). The location $X=X_\tM$ of their maximum $|\psi|$ occurs at the mid-point $X_{\tM 0}=0.5$ at onset but, on increasing $\lambda$ shifts towards the downwelling side boundary due to the quadratic nonlinearity $3\alpha \gS\gS^{\prime}$, exactly as we predict (\ref{R-M-dag-E0}) and find (figures~\ref{fig2}($a$--$d$)) for $R=R_\tM$.


\begin{figure}
\centerline{}
\vskip 3mm
\centerline{
\includegraphics*[width=0.5 \textwidth]{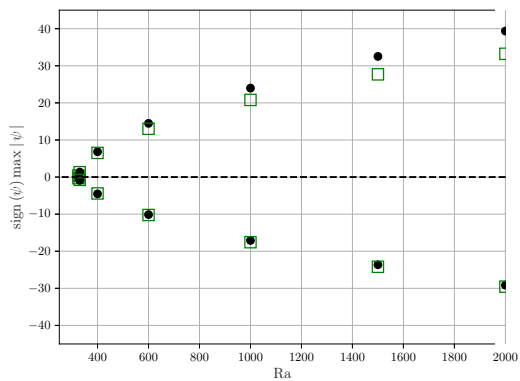}
}
\caption{Bifurcation diagram for the no-rotation case ($E^{-1}=0$), $\sigma=0.1$, showing $\text{sign}(\psi)\text{max}\!\mid\!\psi\!\mid\!$ as a function of $\Ra$. Black dots: DNS, green squares: Asymptotics.}
\label{fig3}
\end{figure}

In figure~\ref{fig3}, we plot the maximum value of $|\psi|$ on the entire domain, but signed depending on whether the solution describes upwelling $\psi>0$ or downwelling $\psi<0$ on the symmetry axis $R=0$. We note that at given $\Ra$ the amplitude max$|\psi|$ of the upwelling solution is greater than that for the downwelling solution. This is a finite amplitude effect that the weakly nonlinear calculation ($\delta\ll 1$) in \S\ref{E-1=0}   could not identify, at any rate to the order taken. Note too that the solution portrayed in figure~\ref{fig2} at $\Ra_c=330$ is very close to the bifurcation point on  figure~\ref{fig3}, yet, as already mentioned, outside the range of validity of our small-$\delta$ weakly nonlinear asymptotics of \S\ref{Small-amplitude-expansion-N}. Bearing those limitations in mind, it is remarkable how well our long length scale (small $\epsilon$) asymptotic amplitude equation (\ref{amp-eq-brief-nd-extra}) works, giving good maximum amplitude up to remarkably large $\Ra=2,000$ and beyond (not portrayed). A partial asymptotic explanation is provided in the second paragraph of \S\ref{lrge-Ra} (below).


\subsection{The rotating case, $E(\,>E_c)$ finite\label{rotating-DNS}}

The  small-$\delta$ analysis of \S\ref{delta-O1} identified a transcritical bifurcation, for which the subcritical branch is presumably unstable. The origin of that instability is encapsulated by the quadratic term $(\beta/\sigma) R^{-1}\gS^2$ in the expression (\ref{amp-eq-brief-nd}$c$) for $\GS$. The analysis of \S\ref{delta-O2}, valid for sufficiently small $\beta/\sigma$, identified possible recovery on a stable steady solution  upper branch. Whether or not such a branch exists for finite $\beta/\sigma$ remains a matter of speculation, a consideration that emphasises the importance of the size of $\beta/\sigma$. From a general point of view, complications that limit the validity of the approach are likely, as $E$ decreases towards $E_c$. Furthermore, the importance of the $(\beta/\sigma)$-term must increase with decreasing $\sigma$. In the following subsections, we investigate how far can we decrease $E$ and $\sigma$ and yet still obtain useful asymptotic results.


\subsubsection{Meridional flow\label{meridional}}


\begin{figure}
\centerline{}
\vskip 3mm
\centerline{
\includegraphics*[width=1.0 \textwidth]{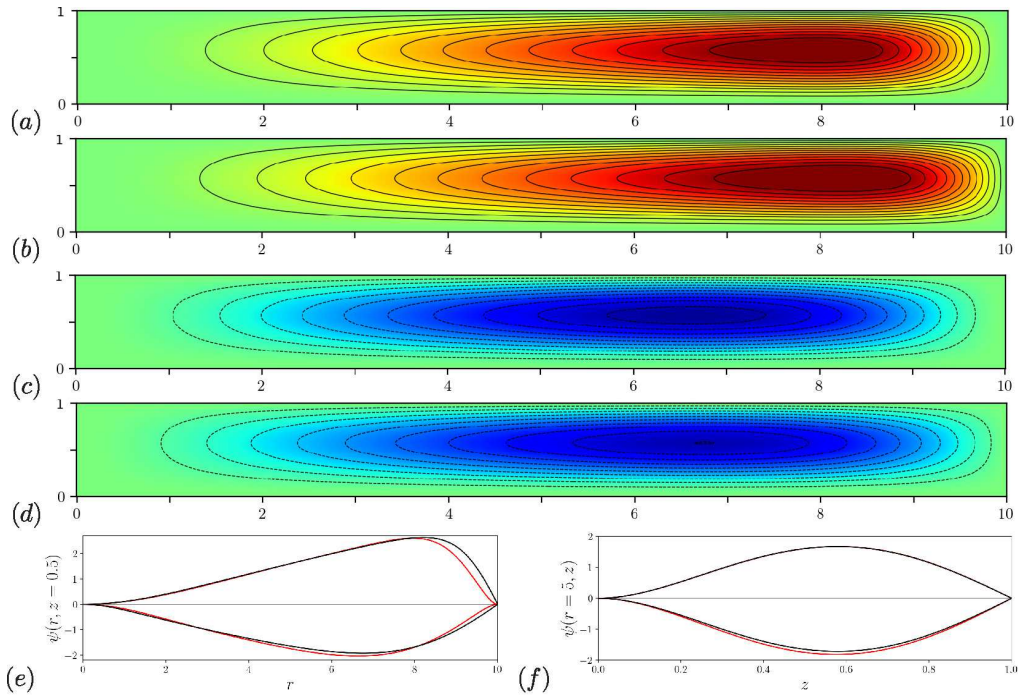}
}
\caption{Rotating case ($E=1$, $\sigma=0.3$, $\Ra=345$). ($a$) $\psi_{\tDNS}>0$; ($b$) $\psi_{\tA}>0$;  ($c$) $\psi_{\tDNS}<0$;  ($d$) $\psi_{\tA}<0$. Colour scale from $0$~(green)~to~$2.5$~(red)~($a$,$b$)  and from $-2.2$~(blue)~to~$0$ ~(green)~($c$,$d$). ($e$) Horizontal cross-sections at $z=0.5$ and ($f$) vertical cross-sections at $r=5$ of the $\psi$ fields shown on ($a$--$d$): DNS ($\psi_{\tDNS}$; red), Asymptotics ($\psi_{\tA}$; black).}
\label{fig4}
\end{figure}



\begin{figure}
\centerline{}
\vskip 3mm
\centerline{
\includegraphics*[width=1.0 \textwidth]{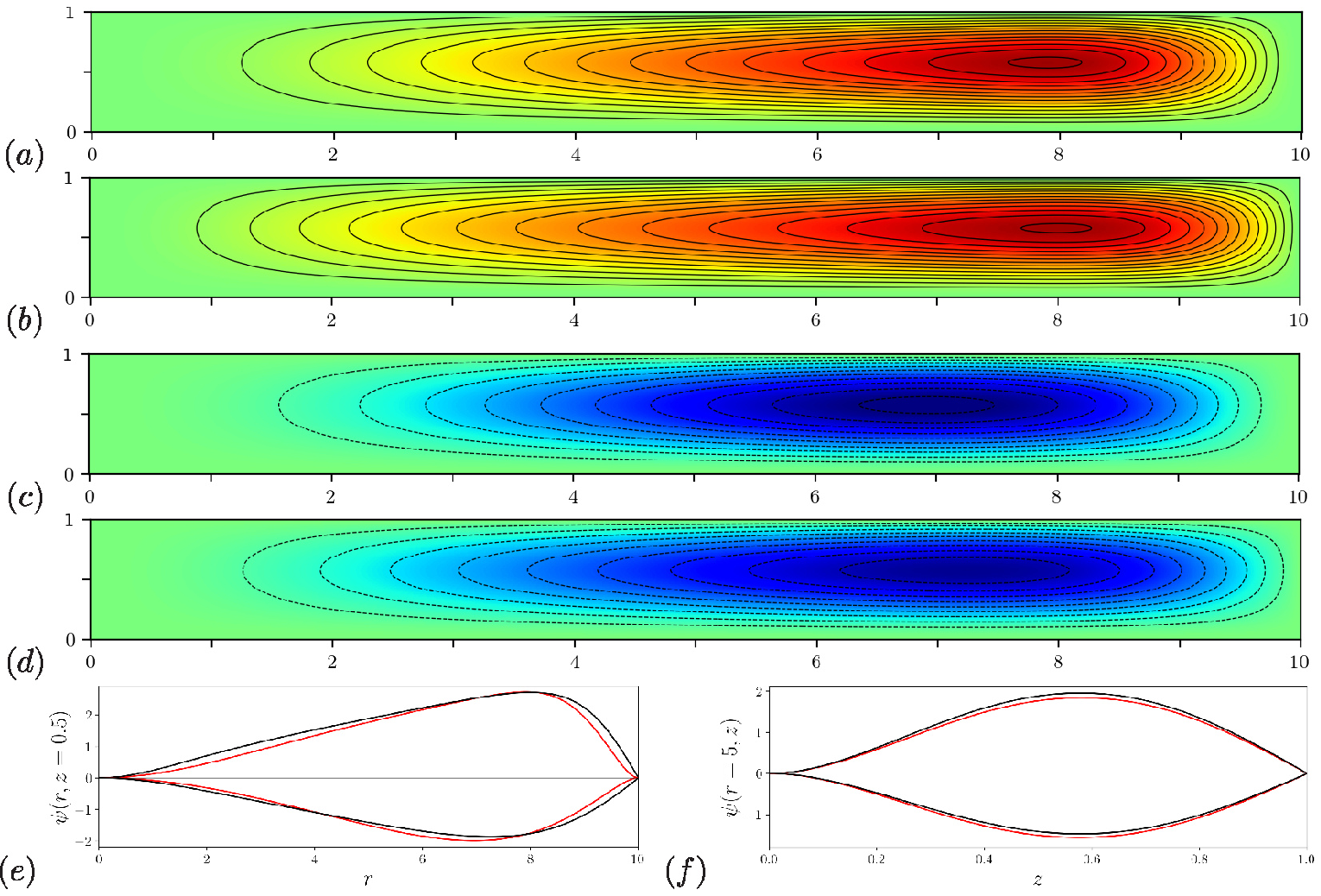}
}
\caption{Rotating case ($E=0.5$, $\sigma=0.3$, $\Ra=360$). $\,\,$
Same as figure~\ref{fig4}. $\,\,$
Colour  scale from~$0$~(green)~to~$3$~(red)~($a$,$b$) and from~$-2$~(blue)~to~$0$~(green)~($c$,$d$).}
\label{fig6}
\end{figure}


Inspection of the asymptotic results illustrated in figure~\ref{fig1} shows that the coefficients (\ref{F-2-coefs*_brief}), which appear in our expression for $\GC_2$ (\ref{F-2_brief-reduced}) of our amplitude equation (\ref{amp-eq-brief}$a$), only vary measurably, on decreasing $E$ from $\infty$, at about $E=1$. On further decrease of $E$, the variation becomes more significant. So, as a tentative first step, we consider the case $E=1$, $\sigma=0.3$ (moderately small), $\Ra=345$. Relative to the critical values $\Ra_c\approx 325.3612$, $\Ra_c^\dag\approx 329.0900$, supercriticality is measured by
\bme
\label{E1-onset}
\be
\aleph\,\approx\,4.26679  \hskip 10mm  \mbox{equivalently}  \hskip 10mm  \lambda \approx 77.3269\,.
\ee
The remaining $\GS$-coefficients are
\be\te
\alpha\,\approx\,0.4868\,\,,\hskip 10mm   \beta\,\approx\,0.1043\,, \hskip 10mm \beta/\sigma\,\approx\,0.34772\,.
\ee
\eme

We illustrate the streamlines for the $\psi\gtrless 0$ in figures~\ref{fig4}($a$--$d$) following the style of figure~\ref{fig2} and exhibiting many of the same features. To highlight any differences between the asymptotics and DNS, we plot horizontal and vertical cross-sections in figures~\ref{fig4}($e$,$f$) respectively. The agreement is almost perfect except for the steep descent curves, $\psi>0$, in figure~\ref{fig4}($e$) between about $r=8$ and the end $r=10$ (recall that $r=10R$). This may be explained by the outer boundary layer caused by the outer rigid boundary condition $\pdr\psi\big|_{r=10}=0$ (\ref{bc_brief-more}$b$), which is not met by the asymptotic solution. A similar, but weaker boundary layer, is evident in the more gently sloping ascent curves, $\psi<0$.


\begin{figure}
\centerline{}
\vskip 3mm
\centerline{
\includegraphics*[width=1.0 \textwidth]{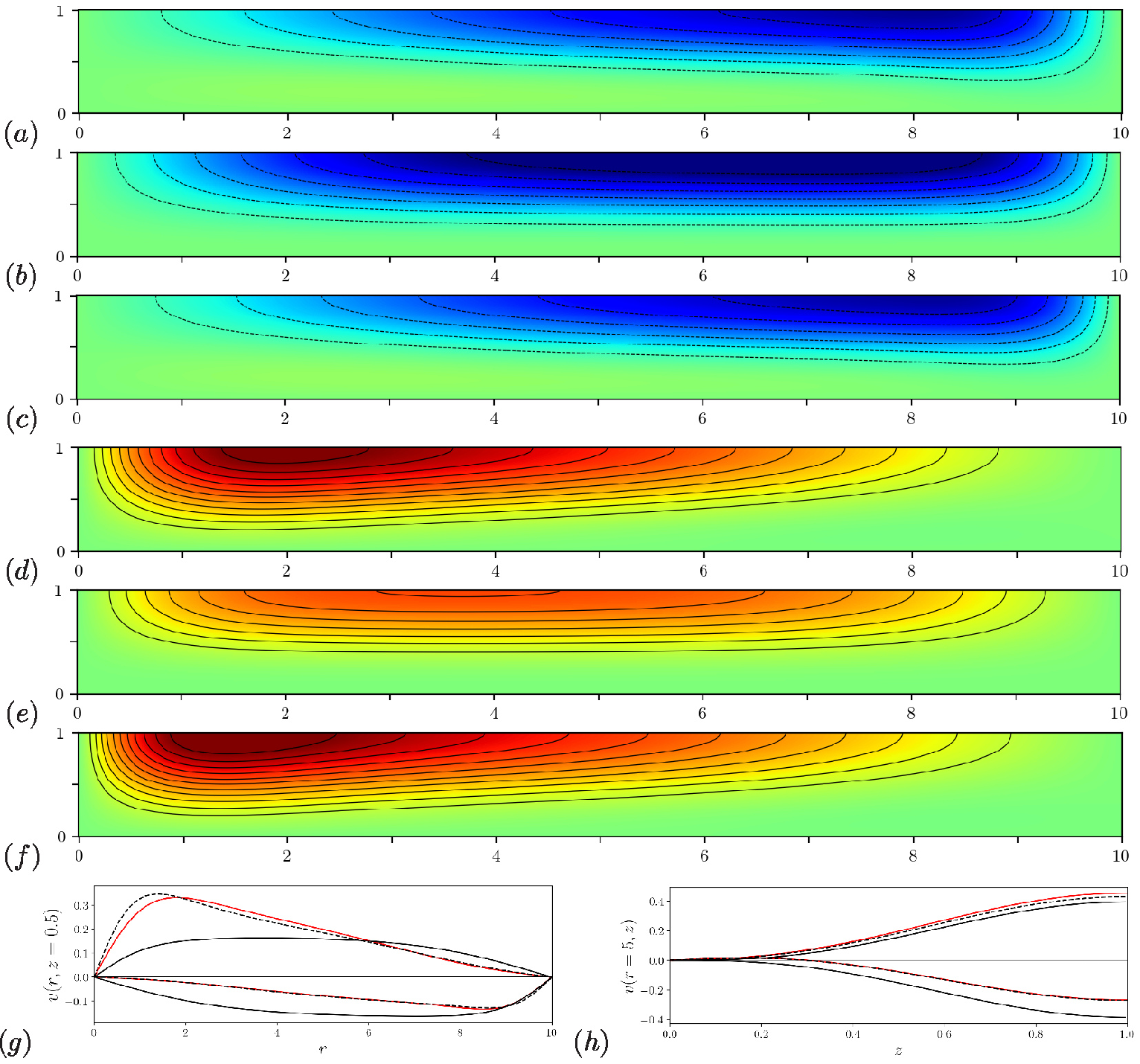}
}
\caption{Rotating case ($E=1$, $\sigma=0.3$, $\Ra=345$). ($a$) $v_{\tDNS}<0$; ($b$) $v_{\tA}<0$; ($c$) $v_{\tH}<0$; ($d$) $v_{\tDNS}>0$;  ($e$) $v_{\tA}>0$;  ($f$) $v_{\tH}>0$. Colour scale from~$-0.35$~(blue)~to~$0_+$ (i.e.~a little positive)~(green)~($a$--$c$) and from~$0$~(green)~to~$0.6$~(red)~($d$--$f$). ($g$) Horizontal cross-sections at $z=0.5$ and ($h$) vertical cross-sections at $r=5$ of the $v$ fields shown on ($a$--$f$): DNS ($v_{\tDNS}$; red), Asymptotics ($v_{\tA}$; solid black), Hybrid ($v_{\tH}$; dashed black).}
\label{fig5}
\end{figure}


Other than the presence of rotation in figure~\ref{fig4}, the use of the lower Prandtl number $\sigma=0.1$ in figure~\ref{fig3} is significant as it increases the influence of inertia. We will return to this point in the following \S\ref{azimuthal}

We test matters further in figure~\ref{fig6}, which addresses the case $E=0.5$ at the same Prandtl number $\sigma=0.3$ but increased Rayleigh number $\Ra=360$. Relative to the critical values $\Ra_c\approx 341.4403$, $\Ra_c^\dag\approx 344.5690 $, supercriticality is measured by
\bme
\label{E0.5-onset}
\be
\aleph\,\approx\,4.95105  \hskip 10mm  \mbox{equivalently}  \hskip 10mm  \lambda \approx 87.3732 \,.
\ee
The remaining $\GS$-coefficients are
\be\te
\alpha\,\approx\,0.2498\,, \hskip 10mm   \beta\,\approx\,0.4453\,, \hskip 10mm \beta/\sigma\,\approx\,1.4843\,.
\ee
\eme
There is not much change in the streamline patterns of figures~\ref{fig6}($a$--$d$) from that displayed in figures~\ref{fig4}($a$--$d$). Indeed, the cross sections in figures~\ref{fig6}($e$,$f$) compare well with similar right-hand boundary layer discrepancies visible in figure~\ref{fig6}($e$). However, a more worrying feature of that figure is the small but clearly evident differences outside that layer between $r=0$ and $8$, which cannot be explained by boundary layer arguments. Indeed studies of even more testing cases of smaller $\sigma$ and/or $E$ reveal even greater $\psi$ differences. For them, the key to the failure is linked to the azimuthal motion due to the rotation.


\begin{figure}
\centerline{}
\vskip 3mm
\centerline{
\includegraphics*[width=1.0 \textwidth]{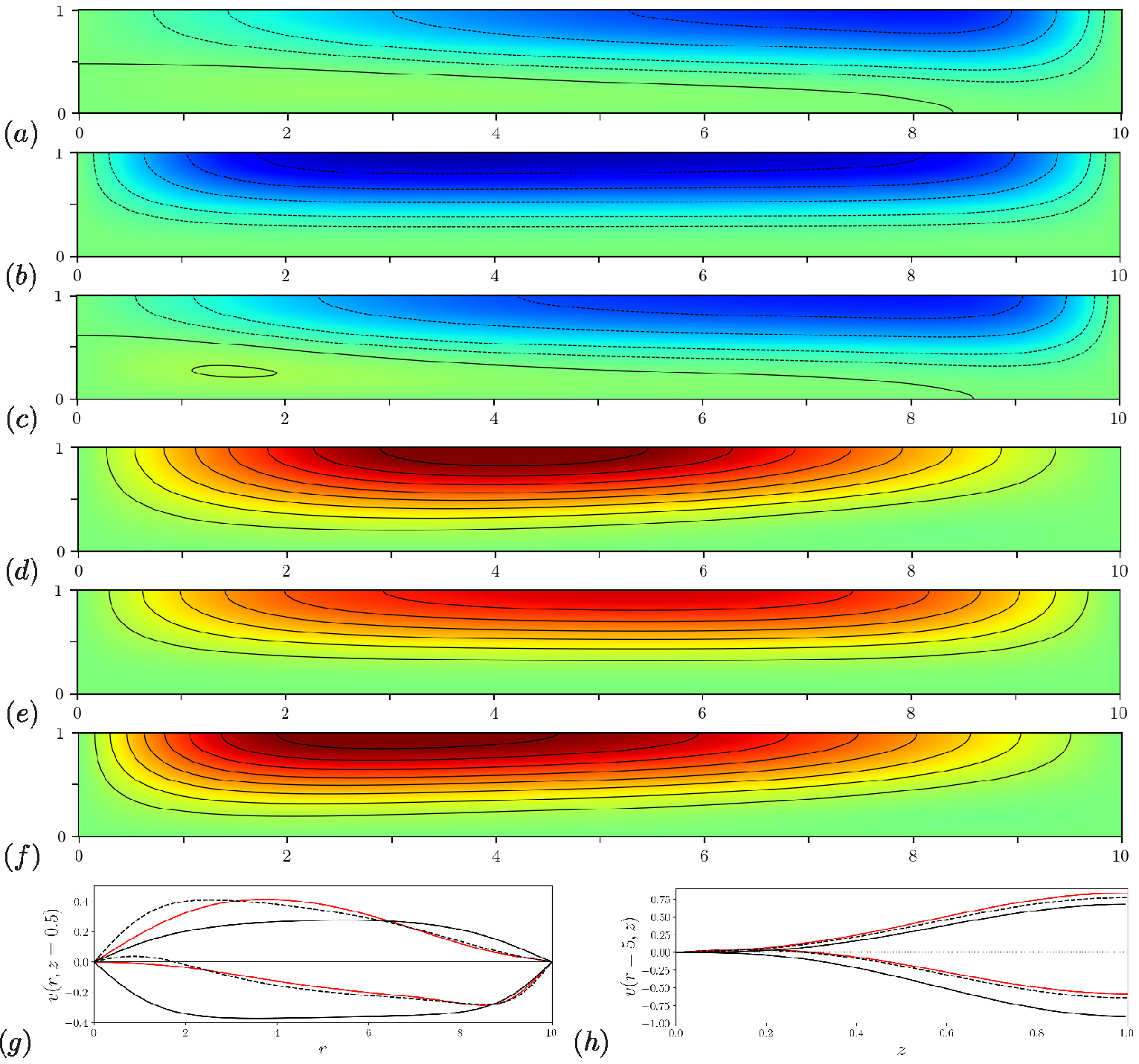}
}
\caption{Rotating case ($E=0.5$, $\sigma=0.3$, $\Ra=360$). $\quad$
Same as figure~\ref{fig5}. $\quad$
Colour scale from~$-1$~(blue)~to~$0_+$ (i.e.~some positive)~(green)~($a$--$c$) and from~$0$~(green)~to~$0.8$~(red)~($d$--$f$).}
\label{fig7}
\end{figure}



\subsubsection{Azimuthal flow\label{azimuthal}}

In \S\ref{meridional} we only considered the meridional flow. The complete solution involves the interaction of the meridional and azimuthal flows through their coupling via the Coriolis force. In this section, we investigate that interaction by considering the azimuthal velocity $v=r\omega$ (see (\ref{psi-varpi}$a$)). We portray the DNS and asymptotic results for $v$ in figures~\ref{fig5}~and~\ref{fig7}, subfigures~($a$,$b$,$d$,$e$), for the cases that correspond to the figures~\ref{fig4}~and~\ref{fig6}, subfigures~($a$,$b$,$c$,$d$), respectively. However, in style, there is one important new addition in figures~\ref{fig5}($c$,$f$)~and~\ref{fig7}($c$,$f$) that we describe as ``Hybrid'', which for the moment must be ignored together with the extra dashed curves on figures~\ref{fig5}($g$,$h$)~and~\ref{fig7}($g$,$h$). On their omission, the remaining comparisons of the DNS and asymptotics are visibly poor. To avoid possible confusion, we stress that, unlike figures~\ref{fig4}($e$)~and~\ref{fig6}($e$), where the upper $\psi>0$ curves correspond to the top figures~\ref{fig4}($a,b$)~and~\ref{fig6}($a,b$), on figures~\ref{fig5}($g$)~and~\ref{fig7}($g$), the lower $v<0$ curves correspond to the top figures~\ref{fig5}($a$--$c$)~and~\ref{fig7}($a$--$c$). Generalised and expressed succinctly, $v\lessgtr 0$ corresponds to $\psi\gtrless 0$ almost everywhere, with some notable exceptions near the axis.

The discrepancies visible in the azimuthal flow contour plots in figures~\ref{fig5}~and~\ref{fig7}, subfigures~($a$,$b$,$d$,$e$) are brought into sharp focus by comparing the red DNS and black asymptotic curves in figures~\ref{fig5}($g$)~and~\ref{fig7}($g$), which describe radial cross sections. Together, they indicate that, for the case of up/downwelling on the axis, the asymptotics over/under estimates the (correctly predicted by the DNS) magnitude $|rv|$ of the angular momentum advected away from/towards the rotation axis in the neighbourhood of the upper boundary. This asymptotic failure is a low  Prandtl number effect, i.e., the increased role of inertia, exacerbated by curvature effects manifested by the various powers of $r$ in the angular momentum equation (\ref{CP2.11-12_brief}$b$), which lead to large azimuthal velocity gradients that violate the long radial length scale assumption on which the asymptotics is based. Though the asymptotic trends are not far off the mark near the outer boundary $r=10$, particularly for the $v<0$ (corresponding to $\psi>0$) curves, they are definitely unsatisfactory elsewhere.

The upshot of the above assessment is that the feedback of the azimuthal flow on the meridional flow is relatively weak in the parameter ranges considered. That said, the azimuthal flow clearly influences the meridional flow as evinced by the fact that the critical Rayleigh number is a function of $E$. So we may suppose that though our azimuthal flow predicted by the asymptotics is flawed, it is sufficiently accurate to generate a totally acceptable meridional flow as illustrated in figures~\ref{fig4}~and~\ref{fig6}.

On the basis that our asymptotically predicted $\psi$ is good, we solved the azimuthal component of the momentum equation, namely (\ref{CP2.11-12_brief}$a$) for the angular velocity $\omega$, with that $\psi$, subject to the $\omega$ boundary conditions appearing in (\ref{bc_brief-more}). We call the solutions of this linear problem ``hybrid'' solutions. The hybrid solutions in figure~\ref{fig5} agree very well with the DNS, vindicating the hybrid approach. On the one hand, it indicates that the asymptotics is on the right track, but that its parameter range of validity is limited. For more testing parameter values, the  hybrid and DNS $v$ shown in figure~\ref{fig7} continue to compare reasonably well but discrepancies are beginning to emerge. They can be explained, from the evidence in figure~\ref{fig6}($e$) that the asymptotic $\psi$-results, on which the hybrid $v$-solution builds, are losing a little accuracy. Evidently, on pushing the parameter values much further, the asymptotic $\psi$ will be too poor to enable the construction of useful hybrid results.

An important feature of the asymptotics is that, at lowest order, the vertical $z$-profile is the same for all $r$, though, of course, the profile amplitude changes. With this restriction, if $v$ has only one sign at some $r$, it cannot exhibit a sign reversal at another~$r$. That means that the DNS and hybrid solutions, portrayed in figures~\ref{fig7}($a$,$c$) exhibiting a sign reversal across the contour beginning on the axis at $z\approx 0.5$ and terminating on the lower boundary just beyond $r=8$, cannot be described by our asymptotics shown in figure~\ref{fig7}($b$).


\subsubsection{A large $\Ra$ application\label{lrge-Ra}}

\begin{figure}
\centerline{}
\vskip 3mm
\centerline{
\includegraphics*[width=1.0 \textwidth]{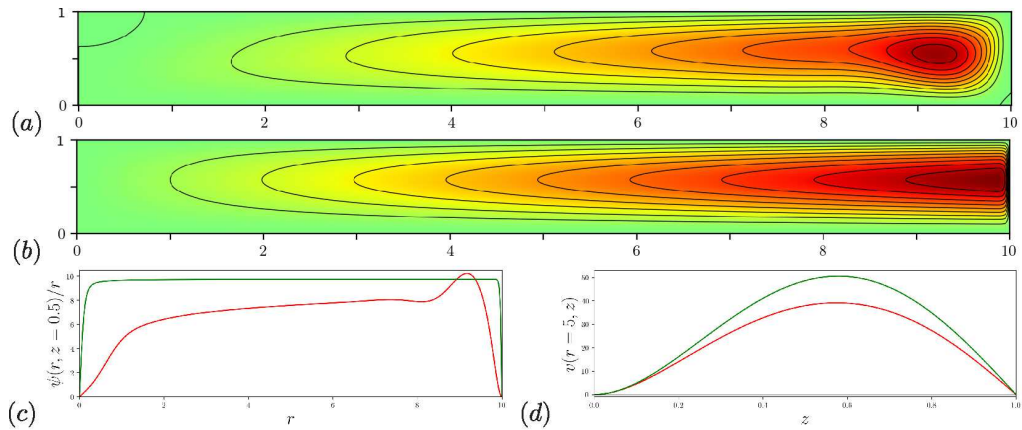}
}
\caption{Rotating case ($E=0.1$, $\sigma=0.5$, $\Ra=15,000$). ($a$) $\psi_{\tDNS}$; ($b$) $\psi_{\tE\infty}$. Colour scale from $-100$  to $100$. ($c$) Horizontal cross-sections at $z=0.5$ and ($d$) vertical cross-sections at $r=5$ of $\psi/r$ for the $\psi$ fields shown on ($a$, $b$): DNS ($\psi_{\tDNS}/r$; red), $E^{-1}=0$ ($\psi_{\tE\infty}/r$; green).
}
\label{fig8}
\end{figure}

\begin{figure}
\centerline{}
\vskip 3mm
\centerline{
\includegraphics*[width=1.0 \textwidth]{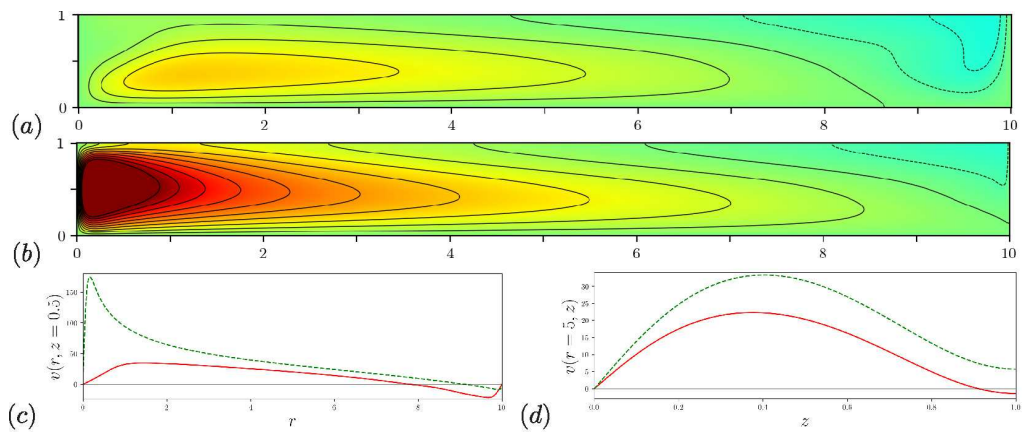} 
}
\caption{Rotating case ($E=0.1$, $\sigma=0.5$, $\Ra=15,000$). ($a$) $v_{\tDNS}$; ($b$) $v_{\tH}$. Colour scale from  $-100$  to  $100$. ($c$) Horizontal cross-sections at $z=0.5$ and ($d$) vertical cross-sections at $r=5$ of the $v$ fields shown on ($a$, $b$): DNS ($v_{\tDNS}$; red), hybrid ($v_{\tH}$; dashed green).
}
\label{fig9}
\end{figure}

\cite{ODD17} portray results for the case $E=0.1$, $\sigma=0.5$, with $\Ra=15000$ in their figures~3($a$--$d$). From our point of view the parameter values, $E=0.1<E_c$ outside the domain of validity for the amplitude modulation equation (\ref{amp-eq-brief}) and $\Ra$ large, are extreme. Nevertheless, it is instructive to make a tentative comparison of the DNS-results, recalculated and displayed as $\psi_\tDNS$ and $v_\tDNS$ in our figures~\ref{fig8} and~\ref{fig9}, with results from a yet more extreme version of our hybrid approach outlined below. Our idea is motivated by the encouraging comparison in figure~\ref{fig3} of our max$|\psi|$-amplitudes for the DNS and asymptotic evaluation of $\psi$ in no-rotation cases at largish $\Ra$. Their robustness suggests that such $\psi=\psi_{\tE\infty}$ (for $E^{-1}=0$) might provide a plausible approximation of $\psi_\tDNS$ at finite rotation (i.e., for $E^{-1}\not=0$); at any rate from a qualitative point of view.

To assess our hypothesis, we plot asymptotic $\psi_{\tE\infty}$--results in figure~\ref{fig8} for the parameter values of \cite{ODD17} but, of course, by definition replace their $E=0.1$ with $E^{-1}=0$. For that case, we recall that $\Ra_c=320$, $\Ra_c^\dag\approx 323.9321$, while, for their large  $\Ra=15000$, we have
\bse
\label{EO-onset-more}
\be
\aleph\,\approx\,3732.3350  \hskip 10mm  \mbox{equivalently}   \hskip 10mm  \lambda \approx54812.7155 \,,
\ee
in place of (\ref{EO-onset}$a$,$b$). Noting that $\beta=0$ and
\be
\alpha\,\approx\, 0.1659 + 0.11848\sigma^{-1}\,\approx\,0.4028 \hskip5mm  \mbox{for} \hskip3mm \sigma=0.5\,,
\ee
in place of (\ref{EO-onset}$c$), the large $\lambda$ asymptotic mainstream solution of (\ref{amp-eq-brief-nd-extra}) is
\be
\gS\,\approx\, \sqrt{\lambda} \,\approx\,234.1212\,.
\ee
\ese
This corresponds to the approximate solution
\bse
\label{amp-eq-brief-psi-P}
\be
0\,=\,\GC_2/g\,\approx\,-\,\mu^2 \Ra_c^{-1} \,+\,\FC_\theta g^2\,=\,-\,\mu^2 \Ra_c^{-1} \,+ \,\Ra_c^2 \bav{P^2}g^2  
\ee
of (\ref{amp-eq-brief}$a$), noting  (\ref{F-2_brief-reduced}$a$) and (\ref{F-2-coefs*_brief}$d$). It describes the balance between the buoyant driving and the nonlinear convection of heat as traced via $R^{-1}\JR\bigl(\psi_0\,,\,\theta_0\,\bigr)= -\Ra_cg^2{\dot P}$ and (\ref{theta-2_z}$a$,$b$). Moreover, (\ref{amp-eq-brief-psi-P}$a$) determines the leading order result
\be
R^{-1}\psi_0\,=\,\Ra_c g P(z)\,\approx\,\mu\Ra_c^{-1/2}P(z)\Big/\bav{P^2}^{1/2}
\ee
(see (\ref{psi-P}$a$)) and, on use of $(\epsilon\mu)^2=\Ra-\Ra_c$ (\ref{Ra-exp}), equivalently
\be
\dfrac{\psi_0}{r}\,\,\approx\,\sqrt{\dfrac{\Ra-\Ra_c}{\Ra_c}}\,\dfrac{P(z)}{\bav{P^2}{}^{1/2}}
\ee
\ese
independent of $r$. The result (\ref{amp-eq-brief-psi-P}$c$) holds everywhere except in the boundary layers, roughly square regions adjacent to the lateral boundaries $r=0$ and $10$, where the solution is invalid. Those layers are evident in figure~\ref{fig8}($c$), which describes the horizontal cross-section $r^{-1}\psi_{\tE\infty}\,(\propto \gS)$ (note the factor $r^{-1}$ absent in previous cross-sections). In the mainstream $2\lessapprox r\lessapprox 8$, the agreement of $r^{-1}\psi_{\tE\infty}$ with $r^{-1}\psi_\tDNS$ for the rotating case, $E=0.1$, is qualitatively remarkable, in view of the tenuous assumptions made. It suggests that rotation modifies but does not control the meridional flow. From this point of view, figure~\ref{fig8} sheds new light on the no-rotation upper branch results portrayed in figure~\ref{fig3}. There, only results up to $\Ra=2,000$ are illustrated, but calculations up to $\Ra=30,000\,(>15,000\mbox{, used in figure~\ref{fig8}})$ were also performed. As the percentage errors ceased to change over that considerable extension, those results are not reported here. The same asymptotic--DNS agreement is also evident on the right of figure~\ref{fig8}($c$), from which the similar sizes of max$\,\psi_{\tE\infty}$ and max$\,\psi_{\tDNS}$ (albeit for $E=0.1$) may be estimated.


We undertook a hybrid calculation (referred to as ``hybrid-$\{E\!=\!\infty\}$''), employing $\psi=\psi_{\tE\infty}$ derived from the $E^{-1}=0$ asymptotics described above, rather than $\psi_\tA$, for which  the asymptotics is irrelevant in the case of interest, $E=0.1$. The $v_\tH$--contours for that hybrid calculation are illustrated on figure~\ref{fig9}($b$). Beyond $r\approx 0.5$, they compare favourably with the DNS illustrated on figure~\ref{fig9}($a$). A more precise quantitative measure comes from the horizontal cross-section in figure~\ref{fig9}($c$). The failure of $v_\tH$, relative to  the true $v_\tDNS$, for $r\lessapprox 5$ is readily traced to the singular behaviour of $r^{-1}\psi_{\tE\infty}\,(\propto \gS)$ for small $r$. The maximum of $v_\tH$ is located at $r\approx 0.3$, which also measures the width of the $\psi_{\tE\infty}$ boundary layer visible near $r=0$ on figure~\ref{fig8}($c$). The presence of strong $v_\tH$ on $0.3\lessapprox r \lessapprox 5$ stems from the overestimation of the strength of the meridional flow, as measured by $\psi_{\tE\infty}$ over that domain. Such strong flow advects the angular momentum $rv_\tH$ and pertains to the proposal in \cite{ODD17,ODD18} that, at large  $\Ra$, angular momentum tends to be constant on streamlines.

At $\Ra=360$, not far above critical, both $v_\tDNS$ and $v_\tH$ portrayed on figures~\ref{fig7}($a$) and~($c$) are generally negative except for a small region close to the lower boundary but terminating before $r=9$, where $v_\tDNS$ and $v_\tH$ are both positive. Interestingly, as $\Ra$ is increased that region of positive $v$ expands to largely fill all the space except for a small region near the outer boundary. This feature is illustrated by $v_\tDNS$ in figure~\ref{fig9}($a$). Remarkably, in view the almost draconian hybrid hypothesis employed, it is also captured by $v_\tH$ in figure~\ref{fig9}($b$). The encouraging agreement vindicates the long horizontal length scale hypothesis for the meridional cell, which is possibly stabilised by the differential rotation caused by angular momentum transport.


\section{Conclusions\label{conclusions}}

Studies of rotating convection \citep[see, e.g.,][and references therein]{Getal14} reveal that large scale vortices are a common feature. They are particularly relevant to atmospheric vortices, such as tropical cyclones and tornadoes. Our asymptotic study has addressed issues raised by the DNS results obtained by \cite{ODD17,ODD18} for axisymmetric convection in a shallow cylinder.

Unlike \cite{Getal14}, who adopted isothermal boundary conditions on the temperature top and bottom, we follow \cite{ODD17} and adopt constant heat flux boundary conditions. This choice is significant, because, for sufficiently large $E$ (small rotation), the onset of instability occurs on a long horizontal length scale. We have taken advantage of this feature and \citep[like][before us]{D88,C98} applied the two-length scale asymptotic approach pioneered by \cite{CP80} for the non-rotating case.

Our investigation of cylindrical geometry highlights effects not apparent in the earlier asymptotic studies, particularly the absence of the $X\mapsto -X$, $\psi \mapsto -\psi$ symmetry which occurs in Cartesian geometry. Even without rotation, that absence is apparent upon comparison of the $E\to\infty$ limit of the heat flux function $\GC_2$ (\ref{F-2_brief-reduced}$a$), containing various powers of $R$, with the Cartesian version (\ref{CP-amp-eq}$b$,$c$) possessing only constant coefficients. Despite this difference, it is encouraging to find that we have no new coefficients and that the \cite{CP80} values (\ref{F-2-coefs-infinity}) also apply to us. Furthermore, instability occurs via a pitchfork bifurcation in both the Cartesian and cylindrical cases. In the latter, one branch corresponds to upwelling on the axis; the other downwelling.

With rotation ($E$ finite), the pitchfork persists in the infinite Prandtl number, $\sigma$, limit. However, on decreasing  $E=\nu/(H^2\Omega)$ and/or $\sigma=\nu/\kappa$, the pitchfork bends at the bifurcation point to reveal locally a transcritical bifurcation that we describe in \S\ref{delta-O1}, with the subcritical upwelling branch unstable and the supercritical downwelling branch stable.  DNS simulations of the complete governing equations suggest that the subcritical branch loses stability but regains stability on the larger amplitude upwelling branch of the bent pitchfork. Indeed, at large enough Rayleigh number $\Ra$, the axial upwelling leads to ``eye'' formation; a region of reversed meridional flow on the axis  \citep[see, e.g.,][figure 5 for  $\Ra=20,000$]{ODD17}. Such features, which have vertical $z$-profiles dependent on $r$, lie outside the scope of our asymptotics, which is based at lowest order on a vertical $z$-profile, independent of $r$.

It is significant that our long horizontal length scale asymptotic requirement is only met at the bifurcation when $E>E_c$ (see (\ref{F-2-coefs-infinity-more}$a$)) and that the small amplitude theory of \S\ref{delta-O2} is only valid for sufficiently large $\sigma$. It means that, on decreasing the value of the kinematic viscosity $\nu$, both $E$ and $\sigma$ decrease in concert with the consequence that the role of inertia, manifest by the Coriolis acceleration or advected momentum, increases.

In the rotating case, within the limitations just described, our asymptotic theory compares well with the DNS at moderate $\Ra$ for $E>E_c$ and $\sigma$ sufficiently large. However, on increasing the vigour of the motion either by increasing $\Ra$ and/or decreasing $\sigma$, the asymptotic theory becomes inadequate. This deficiency does not originate in the meridional momentum equations (see \S\ref{meridional}) but rather in the angular momentum equation. Essentially, the asymptotics cannot cope with the vigorous advection of angular momentum identified in the DNS. To assess this aspect, we adopted hybrid methods in \S\S\ref{azimuthal},~\ref{lrge-Ra}, whereby meridional flows predicted by the asymptotics were employed in DNS simulations of the angular momentum equation alone. The results are illuminating. They culminate in the successful qualitative agreement of the hybrid-$\{E\!=\!\infty\}$ results with the DNS results of \cite{ODD17} portrayed in our figures~\ref{fig8} and~\ref{fig9}, notably for the low Ekman number case $E=0.1<E_c$, outside the range of validity of both the original asymptotic and hybrid methods. It is no surprise to find that angular momentum transfer plays a significant role, as it is essential for the formation of  vortex like structures. The process is magnified on approaching the axis, where it is largely responsible for the discrepancy emerging in the hybrid-$\{E\!=\!\infty\}$ results visible in figures~\ref{fig9}($a$-$c$). That such a long radial length scale meridional cell (see figures~\ref{fig9}($a$-$c$)) is apparently robust even for $E<E_c$ is presumably due to the stabilising role of the differential rotation.

\ \\
{\bf Declaration of Interests.} The authors report no conflict of interest.

\section*{Acknowledgements}

L.O. and E.D. wish to thank the School of Mathematics, Statistics and Physics, Newcastle University for supporting their visit (26--28 March 2022).


\appendix

\section{The solution of the linear problem\label{linear-problem-sol}}

We restate the zeroth order problem for $P$ (\ref{P-problem_brief}) and $W$ (\ref{varpi-W}) in terms of new variables $\UC$ and $\VC$ (also $\RC_c=E^2\Ra_c/ 4$ (\ref{RCc})) defined by the relations
\bme
\label{zero-order_sol-before_brief*}
\begin{align}
P(z)\,=\,&\,\tfrac12 E\bint{\UC}{z}{1}\,,
&\,  W(z)\,=\,&\,\RC_c\bigl(\VC(z)+z-\VC(0)\bigr)\,,\\
{\dot P}(z)\,=\,&\,-\,\tfrac12 E\UC(z)\,, &\,  {\dot W}(z)\,=\,&\,\RC_c\bigl({\dot \VC}(z)+1\bigr)\,,
\end{align}
\eme
where the limits and constants are arranged such that $P(1)=0$ and  $W(0)=0$. We write
\be
\label{zero-order_sol_brief}
\left[\begin{array}{c}  \!\! -\,ER^{-1} \partial\psi_0/ \partial z \!\!\\[0.2em] \!\!R \varpi_0  \end{array}\right]\,=\,2\RC_c
\left[\begin{array}{c}  \!\! \UC(z)  \!\!\\[0.2em] \!\! \VC(z)+z-\VC(0)\!\! \end{array}\right]\,g(R,T)\,,
\ee
so that $\UC(z)$ and $\VC(z)$ satisfy the homogeneous equations
\bme
\label{UV_eq_brief*}
\te\be
{\ddot \VC}\,-\,2\eG^2\UC\,=\,0\,, \hskip 15mm    {\ddot \UC}\,+\,2\eG^2\VC\,=\,0\,,\hskip 15mm  \eG=E^{-1/2}\,.
\ee
\eme
On the one hand, the equivalence of ${\ddot W}+\Ra_c{\dot P}=0$ (the differential of (\ref{varpi-W}$b$)) and (\ref{UV_eq_brief*}$a$) is self-evident.  On the other hand, noting that $W=-\Ra_c\bint{P}{0}{z}$ (\ref{varpi-W}$c$), the integral  ${\dddot {\left.P\right.}}-\RC_c^{-1}W+z=\,$const.~of (\ref{P-problem_brief}$a$) is equivalent to (\ref{UV_eq_brief*}$b$) on identification of the constant of integration with $\VC(0)$, as yet unknown. Together (\ref{varpi-W}$b$) and (\ref{zero-order_sol-before_brief*}$d$) determine
\be
\label{zero-order_sol-more_brief*}
\Ra_c P(z)\,=\,-\,\RC_c\bigl({\dot \VC}(z)+1\bigr).
\ee
Accordingly, the boundary conditions (\ref{P-problem_brief}$b$) become
\be
\label{UV_bc_brief*}
\UC(0)\,=\,{\dot\UC}(1)\,=\,{\dot\VC}(0)\,+\,1\,=\,{\dot\VC}(1)\,+\,1\,=\,0\,.
\ee
Finally, on taking the $z$-average of (\ref{zero-order_sol-more_brief*}), the identity $\Ra_c \oav{P}=-1$ (\ref{P-Ra}$c$) gives
\be
\label{zero-order_sol-more_brief-extra*}
1\,=\,\RC_c\Bigl(\VC(1)-\VC(0) + 1\Bigr)\,\equiv\,\RC_c\Bigl(\bjump{\, \VC\,} +1\Bigr)\,.
\ee

The complex combination
\be
\label{comp-vel*}
\ZC(z)\,=\,\UC(z)\,+\,\iR\VC(z) 
\ee
solves (\ref{UV_eq_brief*}), when
\bme
\label{Zeq*}
\be
{\ddot \ZC}\,-\,2\iR \eG^2\ZC\,=\,0  \hskip 10mm \mbox{equivalently}   \hskip 10mm     E{\ddot \ZC}\,-\,2\iR\ZC\,=\,0 \,.
\ee
\eme
The solution satisfying ${\dot \ZC}(1)=-\iR$ (see (\ref{UV_bc_brief*})) is
\be
\label{Z_sol_brief*}
\ZC(z)\,=\,\ZC(1) \cosh[\eG(1+\iR)(z-1)]\,-\,\tfrac12 \eG^{-1}(1+\iR)\sinh[\eG(1+\iR)(z-1)]\,.
\ee
The application of the remaining boundary conditions $\UC(0)={\dot\VC}(0)+1=0$ at $z=0$ determines the unknown $\ZC(1)=\UC(1)+\iR\VC(1)$ in (\ref{Z_sol_brief*}). After routine but cumbersome calculations, we obtain
\bse
\label{UV-summary_brief*}
\be
\UC(1)\,+\,\iR\VC(1)\,=\,\ZC(1)\,=\,-\,\tfrac12(\eG\Delta)^{-1}\bigl[(\sinh\eG-\sin\eG)^2\,+\,\iR(\cosh\eG-\cos\eG)^2\bigr]
\ee
together with the other remaining $z=0$ values
\begin{align}
\VC(0)\,=\,&\,\tfrac12(\eG\Delta)^{-1}\bigl[(\cosh\eG-\cos\eG)^2+(\sinh\eG+\sin\eG)(\sinh\eG-\sin\eG)\bigr]\,,\\[0.1em]
{\dot \UC}(0)\,=\,&\,\Delta^{-1}(\sinh\eG-\sin\eG)(\cosh\eG-\cos\eG)\,,
\end{align}
where
\be
\Delta\,=\,\tfrac12[\sinh(2\eG)-\sin(2\eG)]\,.
\ee
\ese
 

\section{The $\GC_2$--coefficients\label{G-coefficients}}

The study of the amplitude equation $\pdT{f}=R^{-1} (R\GC_2)^{\prime}$ (\ref{amp-eq-brief}$a$) needs the values of the coefficients (\ref{F-2-coefs*_brief}$d$) and (\ref{F-2-coefs*-more_brief}) that complete the definition of $\GC_2$ (\ref{F-2_brief-reduced}). We now rewrite those coefficients, which are $z$-averages of various combinations of ${\dot P}$ (\ref{zero-order_sol-before_brief*}$c$) and $W$ (\ref{zero-order_sol-before_brief*}$b$), in terms of  $\UC$, $\VC$ instead
\bme
\label{PW-UV*}
\begin{align}
\RC_c^{-1}\oav{W}\,=\,&\,\oav{\VC}+\tfrac12-\VC(0)\,,  &   4E^{-2}\bav{{\dot P}^2}\,=\,&\,\bav{\UC^2}\,,  \\
\hskip 5mm\RC_c^{-1}\bav{zW}\,=\,&\,\bav{z\VC}+\tfrac13-\tfrac12\VC(0)\,, & - 8E^{-3}\bav{{\dot P}^3}\,=\,&\,\bav{\UC^3}\,,\hskip -10mm
\end{align}
\vskip -7mm
\se
\begin{align}
\RC_c^{-2}\bav{W^2}\,=\,&\,\bav{\VC^2}+2\bav{z\VC}-2\VC(0)\oav{\VC}+\VC^2(0)-\VC(0)+\tfrac13\,,\\
-2\RC_c^{-1}E^{-1}\bav{{\dot P}W}\,=\,&\,\bav{\UC\VC}+\bav{z\UC}\,, \\
-2\RC_c^{-2}E^{-1}\bav{{\dot P}W^2}\,=\,&\,\bav{\UC\VC^2}+2\bav{z\UC\VC}-2\VC(0)\oav{\UC\VC}-2\VC(0)\bav{z\UC}+\bav{z^2\UC}\,. \hskip 3mm
\end{align}
\eme
In the following appendix~\ref{Differentials-averages}, we evaluate complex $z$-averages involving $\ZC$ (\ref{comp-vel*}), that embed those in (\ref{PW-UV*}), and simply extract here the needed real and imaginary parts.

From (\ref{ZC-lin-averages_brief*}), the $z$-averages linear in $\UC$ and $\VC$, in addition to $\oav{\UC}=0$, are
\bme 
\label{linear-means_brief*}
\begin{align}
\oav{z\UC}\,=\,&\,-\,\tfrac12 E \RC_c^{-1}\,, &\,  \oav{z^2\UC}\,=\,&\,\tfrac12 E\bigl[-1 -2\VC(1)+ E{\dot \UC}(0)\bigr]\,,\\
\oav{\VC}\,=\,&\,\tfrac12 E {\dot \UC}(0)\,,&\,  \oav{z\VC}\,=\,&\,\tfrac12 E \UC(1)\,.
\end{align}
\eme

From (\ref{ZC-quad-averages_brief*}$b$,$d$,$f$), we may derive the following quadratic $z$-averages
\bme
\label{quad-means_brief*}
\begin{align}
\bav{\UC^2}\,=\,&\,\IC_r\,+\,\JC_r\,+\,\KC\,, &        \bav{\UC\VC}\,=\,&\,\IC_i\,+\,\JC_i\,,\\
\bav{\VC^2}\,=\,&\,-\,\IC_r\,-\,\JC_r\,+\,\KC\,, &     2\bav{z\UC\VC}\,=\,&\,\IC_i\,+\,\HC_i\,.
\end{align}
\eme
Here $4\IC=4(\IC_r+\iR\IC_i)$ (\ref {Wronskian-I_brief*}), evaluated at $z=1$ and $0$, yields the two alternative forms
\bme
\label{quad-means_brief-more*}
\begin{align}
4\IC_r\,=\,&\,\left\{\begin{array}{l}
[\UC(1)]^2\,-\,[\VC(1)]^2\,,\\[0.2em]
  E{\dot \UC}(0)\,-\,\bigl[\VC(0)\bigr]^2\,,
\end{array}\right.   &
4\IC_i\,=\,&\,\left\{\begin{array}{l}
2\UC(1)\VC(1)-\tfrac12 E\,,  \\[0.2em]
\tfrac12E\Bigl[\bigl[{\dot \UC}(0)\bigr]^2-1\Bigr]
\end{array}\right.  
\intertext{respectively (cf.~(\ref{I-int*}$a$,$b$)), while, on writing  $\JC=\JC_r+\iR\JC_i$, $\HC=\HC_r+\iR\HC_i$, (\ref{ZC-quad-averages-coeffs_brief*}) gives}
8E^{-1}\JC_r\,=\,&\,-\UC(1)-\VC(0){\dot \UC}(0)\,,   &  8E^{-1}\JC_i\,=\,&\,1-\RC_c^{-1} \,,\\
4E^{-1}\KC\,=\,&\,-\UC(1)+\VC(0){\dot \UC}(0)\,,   &  8E^{-1} \HC_i\,=\,&\,-2\VC(1)+E{\dot \UC}(0)\,.
\end{align}
\eme
To obtain (\ref{quad-means_brief-more*}$d$), we have noted that, in (\ref{ZC-quad-averages-coeffs_brief*}$a$), $\bjump{\VC}=-1+\RC_c^{-1}$ (see (\ref{zero-order_sol-more_brief-extra*})), while for (\ref{quad-means_brief-more*}$f$), we have noted that, in (\ref{ZC-quad-averages-coeffs_brief*}$c$), $\bjump{\ZC^2}=-\tfrac12 \iR E \bjump{{\dot \ZC}^2}$ (see (\ref{I-int*})) with value (\ref{I-int*}$b$).

The solutions of the simultaneous equations (\ref{UV-cubic-rel*}) determine the only two needed cubic $z$-averages
\bse
\label{UV-cubic-results*}
\begin{align}
  \bav{\UC^3}\,=\,&\,-\dfrac{3}{10}E\bigl[\UC(1)\bigr]^2+\dfrac{1}{30}E^2\bigl[{\dot \UC}(0)\bigr]^3\,=\,\dfrac{E}{30}\UC(1)\bigl[-9\UC(1)+4\VC(1){\dot \UC}(0)\bigr],\\
\bav{\UC\VC^2}\,=\,&\,-\dfrac{1}{10}E\bigl[\UC(1)\bigr]^2+\dfrac{1}{15}E^2\bigl[{\dot \UC}(0)\bigr]^3=\,\dfrac{E}{30}\UC(1)\bigl[-3\UC(1)+8\VC(1){\dot \UC}(0)\bigr].
\end{align}
\ese
  

\section{Differentials and $z$-averages\label{Differentials-averages}}

We consider differentials that we can integrate to determine relations between the various integrals of $\UC$ and $\VC$ appearing in (\ref{PW-UV*}). The essential strategy is to employ the property $E{\ddot \ZC}-2\iR\ZC=0$ (\ref{Zeq*}$b$) to cast intergrands as differentials so that $z$-averages may be integrated with results determined by the end point values of $\ZC$ and ${\dot \ZC}$ at the boundaries $z=0$ and $1$. The useful jump identities
\bme
\label{dotZ-jump*}
\be
\bjump{{\dot \ZC}}\,=\,-\,{\dot \UC}(0)\,, \hskip8mm\mbox{and}\hskip8mm      \bjump{\YC{\dot \ZC}}\,=\,-\,\iR\bjump{\YC}\,-\,\YC(0){\dot \UC}(0)
\ee
\eme
for any complex function $\YC(z)$, follow from the boundary conditions (\ref{UV_bc_brief*}).


\subsection{Linear means\label{lin-int-brief*}}

On making the substitution $E^{-1}\ZC=-\tfrac12 \iR {\ddot \ZC}$ in the left-hand sides of each of the following, integration by parts, possibly aided by (\ref{dotZ-jump*}), yields 
\bse
\label{ZC-lin-averages_brief*}
\begin{align}
E^{-1}\oav{\ZC}\,=\,&\,-\,\tfrac12 \iR \bjump{\dot \ZC}\,=\,\tfrac12 \iR\, {\dot \UC}(0)\,,\\
E^{-1}\oav{z\ZC}\,=\,&\,-\,\tfrac12 \iR \bjump{z{\dot \ZC}-\ZC}\,=\,-\,\tfrac12\Bigl(1\,+\,\bjump{\VC}\Bigr) +\tfrac12 \iR\, \UC(1)\nonumber\\
=\,&\,-\,\tfrac12\RC_c^{-1} +\tfrac12 \iR\, \UC(1)\hskip 20mm \mbox{(use (\ref{zero-order_sol-more_brief-extra*}))}\,,\\
E^{-1}\bav{z^2\ZC}\,=\,&\,-\,\tfrac12 \iR \Bigl(\bjump{z^2{\dot \ZC} -2z\ZC}\,+2 \oav{\ZC}\Bigr)\nonumber\\
=\,&\,\tfrac12\bigl(-1-2\VC(1)+E{\dot \UC}(0)\bigr)+\iR\UC(1)\,.
\end{align}
\ese


\subsection{Quadratic integrals\label{quad-int-brief*}}

Here we take advantage of the Wronskian property
\be
\label{Wronskian-I_brief*}
4\IC\,\equiv\,4\bigl(\IC_r\,+\,\iR\IC_i\bigr)\,=\,\ZC^2+\tfrac12 \iR E{\dot \ZC}^2\,=\,\mbox{complex constant,}
\ee
independent of $z$, i.e., $\dR_z \IC\equiv \dR \IC/\dR z =0$ (use (\ref{Zeq*}$b$)). The trivial consequence $\bjump{\IC}=0$ implies
\bse
\label{I-int*}
\begin{align}
\bjump{\ZC^2}\,=\,&\,\bigl[\UC(1)\bigr]^2\,-\,\bjump{\VC^2}\,+\,2\iR\UC(1)\VC(1)\\
=\,&\,-\,\tfrac12 \iR E \bjump{{\dot \ZC}^2}\,=
\,E \Bigl\{ {\dot \UC}(0) +\tfrac12\iR\bigl[{\dot \UC}(0)\bigr]^2   \Bigr\}\,, \hskip 10mm \mbox{(use (\ref{dotZ-jump*}$b$); $\YC={\dot \ZC}$)\,,}
  \end{align}
a result compatible with the identities (\ref{quad-means_brief-more*}$a$,$b$). Also useful is
\be
\bjump{\ZC\ZC^\ast}\,=\,\bigl[\UC(1)\bigr]^2\,+\,\bjump{\VC^2}\,=\,2 \bigl[\UC(1)\bigr]^2-\,\bjump{\bigl(\ZC^2\bigr)_r}\,,
\ee
\ese
where, as usual, the subscript $r$ denotes the real part.

Our approach is similar to that of appendix~\ref{lin-int-brief*}, but rather than integrate by parts, we proceed directly with the construction of differentials. Accordingly, to establish the identities (\ref{ZC-quad-averages_brief*}$a$,$c$,$e$) below, we perform the differentiation on their right-hand sides and make use of the identity $E{\ddot \ZC}=2\iR\ZC$ (\ref{Zeq*}$b$). Their $z$-averages determine (\ref{ZC-quad-averages_brief*}$b$,$d$,$f$) in
\bme
\label{ZC-quad-averages_brief*}
\begin{align}
\ZC^2\,=\,&\,\dR_z\bigl[2\IC z\,-\,\tfrac14{\iR E}\ZC{\dot \ZC}\bigr]\,,&
\bav{\ZC^2}\,=\,&\,2(\IC\,+\,\JC)\,,\\
\ZC\ZC^\ast\,=\,&\,\dR_z\bigl[\tfrac14{\iR E}\bigl(\ZC{\dot \ZC}^\ast-{\dot \ZC}\ZC^\ast\bigr)\bigr]\,,&
\bav{\ZC\ZC^\ast}\,=\,&\,2 \KC\,,\\
z\ZC^2\,=\,&\,\dR_z\bigl[\IC z^2\,-\,\tfrac14{\iR E}\bigl(z\ZC{\dot \ZC}-\tfrac12\ZC^2\bigr)\bigr]\,,&
\bav{z\ZC^2}\,=\,&\,\IC\,+\,\HC\,,
\end{align}
\eme
where, aided by (\ref{dotZ-jump*}$b$),
\bse
\label{ZC-quad-averages-coeffs_brief*}
\begin{align}
8E^{-1}\JC\,=\,&\,-\,\iR \bjump{\ZC{\dot \ZC}}\,=\,-\,\bigl[\UC(1)+\VC(0){\dot \UC}(0)\bigr]\,-\,\iR\bjump{\VC}\,,\\
4E^{-1}\KC\,=\,&\,\tfrac12 \iR\bjump{\ZC{\dot \ZC}^\ast-{\dot \ZC}\ZC^\ast}\,=\,-\UC(1)+\VC(0){\dot \UC}(0)\,,\\
8E^{-1}\HC\,=\,&\,-\,\iR\bjump{2z\ZC{\dot \ZC}-\ZC^2}\,=\,-2\bigl[\UC(1)+\iR\VC(1)\bigr]\,+\,\iR\bjump{\ZC^2}\,.
\end{align}
\ese


\subsection{Cubic integrals\label{cub-int*}}

We repeat the strategy, leading to (\ref{ZC-quad-averages_brief*}), at the cubic level to construct
\bme
\label{ZC-cubic-averages_brief*}
\begin{align}
3\ZC^3\,=\,&\,-\iR E \,\dR_z\bigl[4\IC {\dot \ZC}\,+\,\tfrac{1}{2}\ZC^2{\dot \ZC}\bigr]\,,&
3\bav{\ZC^3}\,=\,&\,4\iR E\IC{\dot \UC}(0)+4\PC\,,\\
5\ZC^2\ZC^\ast\,=\,&\,\iR E\,\dR_z \bigl[4\IC{\dot \ZC^\ast}-{\dot \ZC}\ZC\ZC^\ast+\tfrac12\ZC^2{\dot \ZC^\ast}\bigr]\,,
\hskip -8mm&&\nonumber\\
&&   5\bav{\ZC^2\ZC^\ast}\,=\,&\,-\,4\iR E\, {\dot \UC}(0)\IC + 4\QC\,,
\end{align}
\eme
in which sequential use of (\ref{quad-means_brief-more*}$b$) and (\ref{quad-means_brief-more*}$a$) yields
\bse
\label{ZC-cubic-averages-coeffs_brief*}
\begin{align}
8\IC_i{\dot\UC}(0)\,=\,&\,E\Bigl[\bigl[{\dot\UC}(0)\bigr]^2-1\Bigr]{\dot\UC}(0)\nonumber\\
\,=\,&\,E\bigl[{\dot\UC}(0)\bigr]^3\,-\,\bigl[\UC(1)\bigr]^2\,+\,\bjump{\VC^2}
\,=\,E\bigl[{\dot\UC}(0)\bigr]^3\,-\,\bjump{\bigl(\ZC^2\bigr)_r}\,,
\intertext{and where, again aided by (\ref{dotZ-jump*}$b$),}
8E^{-1}\PC\,=\,&\,-\,\iR \bjump{\ZC^2{\dot \ZC}}\,=\,-\,\iR  [\VC(0)]^2{\dot \UC}(0)- \bjump{\ZC^2}\,,\\
8E^{-1}\QC\,=\,&\,\iR \bjump{-2{\dot \ZC}\ZC\ZC^\ast+\ZC^2{\dot \ZC^\ast}}\,=\,3\iR[\VC(0)]^2{\dot \UC}(0)-2 \bjump{\ZC\ZC^\ast}- \bjump{\ZC^2}\,.
\end{align}
\ese

Substituting (\ref{ZC-cubic-averages-coeffs_brief*}) into (\ref{ZC-cubic-averages_brief*}$b$,$d$), noting (\ref{I-int*}) and taking real parts yields
\bse
\label{UV-cubic-rel*}
\begin{align}
3\bav{\UC^3}\,-\,9 \bav{\UC\VC^2}\,=\,3\bav{\ZC^3}_r\,=\,&\,-\,\tfrac12 E^2 \bigl[{\dot\UC}(0)\bigr]^3\,,\\
5\bav{\UC^3}\,+\,5 \bav{\UC\VC^2}\,=\,5\bav{\ZC^\ast\ZC^2}_r\,=\,&\,\tfrac12 E^2 \bigl[{\dot\UC}(0)\bigr]^3\,- 2 E[\UC(1)]^2\,.
\end{align}
\ese

\end{document}